\documentclass[12pt,a4]{article}
\usepackage[utf8]{inputenc}
\usepackage{graphicx}
\usepackage[normalem]{ulem}
\usepackage{amsthm,amsmath,latexsym,amssymb,amsfonts,amscd}
\usepackage{graphics,lscape,fancyhdr,array,stmaryrd,euscript,color}
\usepackage{ifpdf}
\usepackage{verbatim,slashed}
\usepackage{relsize,setspace}
\numberwithin{equation}{section}
\usepackage{hyperref}

\renewcommand{\tilde}{\widetilde}

\newcommand{\bref}[1]{\textbf{\ref{#1}}}

\newcommand{\im}{\mathop{\mathrm{Im}}}
\newcommand{\dd}{\partial}
\renewcommand{\d}{\partial}

\newcommand{\binner}[2]{%
  {\langle}\kern-4.15pt{\langle}#1{,}\,#2{\rangle}\kern-4.15pt{\rangle}}

\newcommand{\qcommut}[2]{[#1{,}\,#2]_\star}
\newcommand{\pb}[2]{\left\{{}#1{},{}#2{}\right\}}

\newcommand{\half}{\mathchoice{%
    \ffrac{1}{2}}{\frac{1}{2}}{\frac{1}{2}}{\frac{1}{2}}}

\newcommand{\ffrac}[2]{\raisebox{.5pt}%
  {\footnotesize$\displaystyle\frac{#1}{#2}$}\kern1pt}

\newcommand{\dl}[1]{\mathchoice{\ffrac{\dd}{\dd #1}}{\frac{\dd}{\dd
      #1}}{\ffrac{\dd}{\dd #1}}{\ffrac{\dd}{\dd #1}}}

\newcommand{\ddl}[2]{\ffrac{\dd #1}{\dd #2}}

\def\BG-Poincare{Barnich:2009jy}
\def\Fedosov-book{Fedosov:1996fu}

\newcommand{\sdd}{S^\dagger}
\newcommand{\besubeqs}{\begin{subequations}}
\newcommand{\esubeqs}{\end{subequations}}

 \usepackage[numbers,sort&compress]{natbib}
 \usepackage[nottoc]{tocbibind}

\newcommand{\pl}{\partial}

\begin{document}
%\vspace*{1cm}
\begin{center}
{\Large\bfseries 
Higher Spin Extension \\ 
\vspace{4pt}
of Fefferman-Graham Construction 
\vspace{0.4cm}
} \\

\vskip 0.04\textheight

Xavier Bekaert${}^{a}$, Maxim Grigoriev${}^{c}$ and Evgeny Skvortsov${}^{b,c}$

\vskip 0.04\textheight

{\em$^a$ Laboratoire de Math\'ematiques et Physique Th\'eorique\\
Unit\'e Mixte de Recherche $7350$ du CNRS\\
F\'ed\'eration de Recherche $2964$ Denis Poisson\\
Universit\'e Fran\c{c}ois Rabelais, Parc de Grandmont\\
37200 Tours, France}

\vspace{5pt}
{\em$^b$  Albert Einstein Institute, \\
Am M\"{u}hlenberg 1, D-14476, Potsdam-Golm, Germany}

\vspace{5pt}
{\em$^c$ Lebedev Institute of Physics, \\
Leninsky ave. 53, 119991 Moscow, Russia}\\

\vskip 0.02\textheight

{\bf Abstract }

\end{center}
\begin{quotation}
Fefferman-Graham ambient construction can be formulated as $\mathfrak{sp}(2)$-algebra relations on three Hamiltonian constraint functions on ambient  space. This formulation admits a simple extension that leads to higher-spin fields, both conformal gauge fields and usual massless fields on anti-de Sitter spacetime. For the bulk version of the system, we study its possible on-shell version which is formally consistent and reproduces conformal higher-spin fields on the boundary.
Interpretation of the proposed on-shell version crucially depends 
on the choice of the functional class. Although the choice leading to fully interacting higher-spin theory in the bulk is not known, we demonstrate that the system has a vacuum solution describing general higher-spin flat backgrounds. Moreover, we propose a functional class such that the system describes propagation of higher-spin fields over any higher-spin flat background, reproducing all the structures that determine the known nonlinear higher-spin equations. 
\end{quotation}

\pagestyle{empty}

\newpage

\setcounter{page}{1}

\pagestyle{plain}

\tableofcontents

\newpage
%%%%%%%%%%%%%%%%%%%%%%%%%%%%%%%%%%%%%%%%%%%%%%%%%%%%%%%%%%%%%%%%
\section{Introduction}
%%%%%%%%%%%%%%%%%%%%%%%%%%%%%%%%%%%%%%%%%%%%%%%%%%%%%%%%%%%%%%%%
The theory of  higher-spin gravity is intimately tied to Anti de Sitter / Conformal Field theory (AdS/CFT) correspondence \cite{Maldacena:1997re,Witten:1998qj,Gubser:1998bc} in the exotic regime of strong curvature / weak coupling \cite{Sundborg:2000wp,Sezgin:2002rt,Klebanov:2002ja}. Historically, the discovery of the deep relationship between AdS massless fields and elementary fields living on the conformal boundary (aka ``singletons'')  by Flato and Fronsdal \cite{Flato:1978qz} anticipated some ideas of AdS/CFT correspondence. Most presumably, both subjects (higher spins and holography) still hold important insights worth exploring for their mutual benefit. Particular examples where the connection between both subjects might deserve to be explored further are the ambient construction of Fefferman and Graham and its relation to effective actions and higher-spin gauge fields that we investigate in this work.

First of all, the Fefferman-Graham (FG) ambient construction\footnote{The seminal paper is \cite{FG} but see e.g. \cite{Fefferman:2007rka} (and refs therein) for a comprehensive overview on the subject.} is one of the most important mathematical pillar sustaining the AdS/CFT correspondence since its very birth. In fact, it was instrumental in the holographic prescription, see e.g.  \cite{Witten:1998qj,deHaro:2000vlm,Henningson:1998gx}. In its  simplest version it amounts to the flat ambient space approach~\cite{Dirac:1935zz,Dirac:1936fq} whose underlying idea  is to make conformal and/or AdS symmetries manifest: conformal algebra acts on the projective hypercone, while AdS algebra acts on the hyperboloid.

The FG construction can be understood as the curved generalization of the naive ambient space approach. It can be used in several different ways: 
to study curved conformal geometry with the tools of the Riemannian geometry by extending the conformal structure off the hypercone; to study Einstein equations with cosmological constant on the hyperboloid in terms of the Ricci flat ambient space geometry. Also, for odd bulk dimensions the conformal gravity equations on the boundary arise as an obstruction to extending the conformal structure to the FG ambient metric and can also be seen as arising from holographic Weyl anomaly~\cite{Henningson:1998gx}. In a similar way, equations of motion/Lagrangians of conformal gauge fields arise as an obstruction/holographic anomaly of the respective AdS gauge fields~\cite{Henningson:1998gx,Liu:1998bu,Metsaev:2009ym,Bekaert:2012vt,Bekaert:2013zya}.

Typically, it is assumed that in the FG construction the ambient metric is subject to Ricci flatness condition. With this condition omitted the FG construction is equivalent to Hamiltonian constraints: three functions on ambient phase space (with Poisson bracket $\{X^A,P_B\}=\delta^A_B$) that obey the $\mathfrak{sp}(2)$ algebra,\footnote{The fact that there is a triplet of operators that forms $\mathfrak{sp}(2)$ and it is closely related to the FG construction was pointed out in \cite{GJMS} (see also \cite{Bars:2000qm,Bonezzi:2010jr,Bonezzi:2014nua,Bonezzi:2015bza}). In the context of quantizing particle models this system also appeared in \cite{Marnelius:1978fs}.} 
\begin{equation}
 \pb{F_+}{F_-}=F_0 \,, \qquad \pb{F_0}{F_\pm}=\pm2\,F_\pm\,.
 %\quad \pb{F_2}{F_1}=-2F_1
\end{equation} 
and is referred to as ``off-shell FG construction'' in what follows.
These three constraints have a specific form: they contain no more than two powers of momenta. More precisely, one constraint is independent of the momenta, $F_-$, one is linear, $F_0$, and one is quadratic, $F_+$. The constraints can be shown to imply the existence of an ambient metric $G_{AB}(X)$ and a homothety vector field (closely related to what is known as ``compensator field'' in physics literature) $V^A(X)$ that satisfy the FG conditions, which can be summarized as the relation $G_{AB}=\nabla_A V_B$.

When the off-shell FG construction is realized as $\mathfrak{sp}(2)$ constraints, a higher-spin (HS) extension is naturally obtained by allowing the constraints to be arbitrary functions on the phase space.\footnote{The $\mathfrak{sp}(2)$ algebra (or its extensions) plays a prominent role in the unfolded approach \cite{Vasiliev:1988sa} to HS theory at the nonlinear level \cite{Vasiliev:1999ba,Sezgin:2001ij,Vasiliev:2003ev}, but within a rather different framework.} Since the HS extension is done in the ambient space, one can consider it either in the vicinity of the (curved) hypercone $V^2=0$ or in the vicinity of the (curved) hyperboloid $V^2=-1$, which leads to two interpretations in terms of HS gauge fields on the conformal space (the projectivization of the hypercone) and HS fields on  the hyperboloid. 

The formulations of HS fields based on the ambient space $\mathfrak{sp}(2)$-system have originally appeared in the literature independently of FG ambient construction. The idea to describe a tower of HS fields on projective hypercone as an $\mathfrak{sp}(2)$-system in ambient space was proposed in \cite{Bars:2001um} and developed further in \cite{Bonezzi:2010jr,Bonezzi:2014nua}. In the context of HS fields on AdS an $\mathfrak{sp}(2)$-system was proposed in~\cite{Grigoriev:2012xg} where it was shown to describe off-shell HS fields upon linearizing over the AdS vacuum solution. The same system describes~\cite{Bekaert:2013zya,Alkalaev:2014nsa} the off-shell theory of conformal higher-spin (CHS) fields \cite{Fradkin:1985am,Segal:2002gd,Tseytlin:2002gz,Bekaert:2010ky} provided one subjects the system to certain extra algebraic gauge symmetries. 

CHS fields arise as natural sources for conserved higher-rank tensors, likewise the conformal graviton is a source for the stress-tensor in a CFT. Infinite multiplets of CHS fields are sources for HS currents that are present in free CFT's \cite{Maldacena:2011jn,Alba:2013yda,Boulanger:2013zza,Stanev:2013qra,Alba:2015upa}. When CHS fields are viewed as infinitesimal sources, the effective action is simply a generating functional of the free CFT correlators. While it is trivial to couple the free scalar field to an arbitrary gravitational background, a remarkable fact \cite{Segal:2002gd} is that one can couple the free scalar field to an arbitrary CHS background as well, i.e. to extend CHS fields beyond infinitesimal sources. This requires an intricate structure of non-abelian symmetries that make the effective action of the free scalar field gauge-invariant on an arbitrary CHS background. It is these gauge symmetries for the CHS sources that the $\mathfrak{sp}(2)$-system describes,  thereby encoding information about the effective action on any background. Moreover, for even boundary dimension the effective action has a local log-divergent part: it is the conformal gravity action if the background is gravitational and the action of CHS gravity if a HS background is turned on \cite{Fradkin:1985am,Tseytlin:2002gz,Segal:2002gd,Bekaert:2010ky}.

On the hyperboloid, $V^2=-1$, the same $\mathfrak{sp}(2)$-system describes off-shell nonlinear bulk HS gauge fields. More precisely, the $\mathfrak{sp}(2)$-system linearized over the AdS background can be put on-shell, giving the ambient space description of the Fronsdal fields~\cite{Fronsdal:1978rb}. However, it is not clear how to extend this beyond the free approximation: a natural suggestion to put the system on-shell at higher orders is to introduce extra gauge symmetry factoring out the ideal generated by the $\mathfrak{sp}(2)$-fields themselves. This extra gauge symmetry is precisely the one needed to describe CHS fields on the boundary and can also be seen as a natural gauge symmetry of the constrained system with constraints $F_i$, which is related to a redefinition of the constraints.\footnote{A somewhat similar factorization procedure is employed in the Vasiliev system~\cite{Vasiliev:2003ev} in general space-time dimension. Precisely this gauge symmetry and its constrained system interpretation was proposed in~\cite{Grigoriev:2006tt}.}

On general grounds, putting the system on-shell by
gauging away off-shell modes crucially depends on the choice of the functional class. For instance, with naive but natural choice the procedure yields an empty system. When formulated in such terms the problem seems to be closely related to the issue of locality in field theory, more specifically in higher-spin theories.\footnote{The degree of non-locality was quantified in \cite{Bekaert:2015tva,Sleight:2017pcz,Ponomarev:2017qab} by reconstructing the quartic vertex, which revealed that the vertex is highly non-local. It should always be possible to manufacture interactions in AdS that would give the expected CFT correlation functions \cite{Bekaert:2014cea,Kessel:2015kna,Bekaert:2015tva,Skvortsov:2015pea,Sleight:2016dba}, but it is unclear how to fix such interactions in the bulk without having to invoke the AdS/CFT argument, which is due to a high degree of non-locality \cite{Sleight:2017pcz,Ponomarev:2017qab}, the problem being similar to that in flat spacetime. }  Indeed, the choice of a functional class controls the derivative expansion of interactions, which is always strongly-coupled in higher-spin theories due to the dimensionless coupling constant and unbounded number of derivatives starting from the quartic order, see e.g. discussion in \cite{Bekaert:2010hw}.

Instrumental in investigating various properties of gauge theories (in particular the $\mathfrak{sp}(2)$-system we are interested in) is the parent approach \cite{Barnich:2004cr,Barnich:2006pc,Barnich:2010sw,Grigoriev:2010ic}. One of the advantages of the parent approach in the context of the AdS/CFT correspondence is that one can jump directly between bulk and boundary simply by changing the compensator field from timelike ($V^2=-1$) to null ($V^2=0$)~\cite{Bekaert:2012vt,Bekaert:2013zya}. The parent equations of motion then rearrange themselves in accordance with the representation structure of the AdS/conformal algebra. 

In this work, with the help of the parent approach we demonstrate that the $\mathfrak{sp}(2)$-system can be pushed one step further: the system has a class of exact solutions --- higher-spin flat backgrounds. We also show that one can put the HS fields on-shell over a HS-flat background, which is not necessarily AdS and thereby probes interactions. More specifically, we propose a suitable functional class in the auxiliary space of the parent formulation, which allows one to put the system on-shell. The resulting equations have the correct form of a flatness condition deformed by a two-cocycle of the HS algebra~\cite{Sharapov:2017yde}. These are the data that completely determine the Vasiliev equations \cite{Vasiliev:1988sa,Vasiliev:2003ev}.  

It has been known for decades that HS fields are hard to make propagate consistently on anything but constant curvature backgrounds \cite{Aragone:1979hx}. Nevertheless, we observe that special backgrounds, those given by flat connections of HS algebras, allow for propagation of HS fields. The flatness condition is hard to interpret from the vantage point of Fronsdal fields \cite{Campoleoni:2012hp}. It is worth mentioning that HS-flat backgrounds were shown to describe rich physics of HS black holes in three-dimensions, see e.g. \cite{Ammon:2012wc} and references therein/thereon.

The outline of the paper is as follows. In section \bref{offshellFG}, we discuss the off-shell FG construction and show that it is equivalent to the $\mathfrak{sp}(2)$ constraints (with some details delegated to appendix \bref{hyporth}). In section \bref{onshellFG}, we review and discuss the on-shell FG construction. In section \bref{sec:hsextension}, a HS extension is proposed and it is discussed how it is related to the known HS systems. In section \bref{sec:hsonshell}, we show that the HS extension can describe fluctuations of massless HS fields over any HS-flat background. Conclusions and discussion are in section \bref{sec:discussion}.

%%%%%%%%%%%%%%%%%%%%%%%%%%%%%%%%%%%%%%%%%%%%%%%%%%%%%%%%%%%%%%%%
\section{Off-shell Fefferman-Graham Theory}\label{offshellFG}
%%%%%%%%%%%%%%%%%%%%%%%%%%%%%%%%%%%%%%%%%%%%%%%%%%%%%%%%%%%%%%%%

By ``off-shell gravity'' in $d+1$ dimensions, we understand a gauge theory whose fields are the components of the metric tensor $g_{\mu\nu}$ ($\mu,\nu=0,1,\ldots,d$), which is assumed invertible and only\footnote{i.e. it is ``off-shell'' in the sense that the fields are \textit{not} subject to field equations.}
subject to the usual gauge transformations (infinitesimal diffeomorphisms):
\begin{equation}
\delta_{\xi} g_{\mu\nu}={\cal L}_\xi g_{\mu\nu}:= \xi^\rho \d_\rho g_{\mu\nu}+\d_\mu \xi^\rho  g_{\rho\nu}+\d_\nu\xi^\rho g_{\mu\rho}\,,
\end{equation}
where $\xi^\mu$ are the components of a gauge parameter which is assumed unconstrained.

Similarly one defines ``off-shell conformal gravity'' in $d$ dimensions via an invertible metric tensor $g_{ab}$ ($a,b=0,1,\ldots,d-1$) by introducing the extra gauge transformations (infinitesimal Weyl transformations): 
 \begin{equation}
 \delta_{\xi,\,\omega} g_{ab}={\cal L}_\xi g_{ab}
 %= \xi^c \d_c g_{ab}+\d_a\xi^c  g_{cb}+\d_b\xi^c g_{ac}
 +2\,\omega\, g_{ab}\,,
\end{equation} 
where $\xi^a$ and $\omega$ are, respectively, parameters of the infinitesimal diffeomorphisms and Weyl transformations.

The ``off-shell FG theory'' defined in ($d+2$)-dimensional ambient space (see Appendix \bref{app:xav} for historical overview and technical details) parameterized by the coordinates $X^A$  requires two ingredients: a nondegenerate ambient metric $G_{AB}$ ($A,B=0,0',1,\ldots,d$) and a nowhere-vanishing homothety vector field $V^A$ that are assumed to obey 
\begin{align}
    \mathcal{L}_V G_{AB}&=2\,G_{AB}\,, & \pl_AV_B-\pl_B V_A=0\,,&
\end{align}
from which it follows that $V_A=\pl_A (V^2/2)$ and $G_{AB}=\nabla_A V_B$. The fields $G_{AB}$ and $V_A$ are subject to usual diffeomorphisms as gauge transformations.

Equivalently, as is shown in appendix \bref{hyporth}, the same relations can be defined in terms of three totally symmetric polyvector fields of rank 2, 1 and 0\,: $G^{AB}(X)$, $V^A(X)$, and $F(X)$ defined on the ($d+2$)-dimensional
ambient space. The equations of motion can be reformulated as the $\mathfrak{sp}(2)$ algebra relations:
\begin{equation}
\label{sp2-rel}
 \pb{F_+}{F_-}=F_0 \,, \qquad \pb{F_0}{F_\pm}=\pm2\,F_\pm\,, 
 %\quad \pb{F_2}{F_1}=-2F_1
\end{equation} 
where
\begin{equation}\label{sp2constr}
 F_+(X,P)=\half\, G^{AB}(X)\,P_AP_B\,, \qquad F_0(X,P)=V^A(X)\, P_A\,, 
 \qquad F_-(X,P)=F(X)\,,
\end{equation}
i.e. we encoded the tensor fields in the three generating functions $F_i$ \,($i=+,-,0$) using extra variables $P_A$ which are ambient momenta, conjugate to the coordinates $X^A$. The Poisson bracket $\pb{\,}{}$ is defined by
\begin{equation}
 \pb{X^B}{P_A}=\delta_A^B\,.
\end{equation}
The gauge symmetries in the FG ambient theory are given by
\begin{equation}
\label{gauge-sym}
 \delta_\xi F_i=\pb{\xi}{F_i}\,,
\end{equation} 
where $\xi=\xi^A(X)\,P_A$ is the generating function of the gauge parameters. It is clear that these gauge transformations are nothing but infinitesimal diffeomorphisms of the ambient space. At the same time, these are particular canonical transformations of the phase space $X,P$.

As explained in appendix \bref{hyporth}, when $G^{AB}$ is nondegenerate (and can thus be seen as the inverse of a metric $G_{AB}$), it follows from the three equations \eqref{sp2-rel} on the three functions \eqref{sp2constr} that 
\begin{equation}\label{sp23}
F(X)= -\half G_{AB}(X) V^A(X) V^B(X)
\end{equation}
and
\begin{equation}\label{FGsp2}
%{\cal L}_V G_{AB}=2\,G_{AB}\,, \qquad 
\nabla_A V^B=\delta_A^B\,,
\end{equation} 
where $\nabla$ is the Levi-Civita connection of the ambient metric $G_{AB}(X)$.

The off-shell FG theory in $d+2$ dimensions is equivalent to off-shell gravity in $d+1$ dimensions provided one disregards the direction along $V^A(x)$ as a genuine spacetime dimension. 
Other way around, any metric $g_{\mu\nu}$ in $d+1$ dimensions can be lifted to an ambient metric $G_{AB}$ and homothety vector field $V^A$ defined on the ($d+2$)-dimensional ambient space such that the original space is a ``curved hyperboloid'' determined by $G_{AB}V^AV^B=-1$,
while the original metric is a pullback of $G_{AB}$ to this level surface.

%%%%%%%%%%%%%%%%%%%%%%%%%%%%%%%%%%%%%%%%%%%%%%%%%%%%%%%%%%%%%%%%
\section{On-shell Fefferman-Graham Theory}
\label{onshellFG}
%%%%%%%%%%%%%%%%%%%%%%%%%%%%%%%%%%%%%%%%%%%%%%%%%%%%%%%%%%%%%%%%
We begin with the $\mathfrak{sp}(2)$ system, i.e. \eqref{sp2-rel} and \eqref{gauge-sym}:
\begin{align}
\label{sp2eqA}
 \pb{F_i\,}{F_j}&=C^k_{ij}F_k\,, &&\delta_\epsilon F_i=\pb{\epsilon}{F_i}\,,
\end{align} 
which we write in a compact way using $\mathfrak{sp}(2)$ structure constants $C^k_{ij}$. Following~\cite{FG}, let us impose an extra condition on the ambient metric entering $F_\pm$: 
\begin{align}\label{Ricci}
    %Ric(G)=0
    R_{AB}=0\,,
\end{align}
i.e. one requires the ambient metric $G_{AB}$ to be Ricci flat.
The system \eqref{sp2eqA}-\eqref{Ricci} defines ``on-shell FG theory''. More precisely, the ambient system should be understood within a certain expansion scheme, called the FG expansion.

There are two interpretations of the on-shell FG theory in $d+2$ dimensions:
\begin{itemize}
    \item This system is equivalent to on-shell gravity in $d+1$ dimensions with a nonvanishing cosmological constant (in other words the metric $g_{\mu\nu}$ is Einstein). The spacetime manifold can be identified with the curved hyperboloid $V^2=-1$;
    
    \item This system describes conformal gravity in $d$ dimensions. For $d$ odd it is off-shell, while for $d$ even it is on-shell, the field equations resulting from the conformal anomaly.\footnote{For $d$ even, in the original FG approach the Ricci flatness was imposed only up to a certain power of the defining function so that the conformal gravity was always off-shell. Another point of view is to require Ricci flatness at all orders which results in conformal gravity equations. Note that in this case the system also describes subleading solutions.} The spacetime manifold can be identified with the projectivization of the curved hypercone $V^2=0$;
\end{itemize}

There is another way to describe off-shell conformal gravity by introducing in place of~\eqref{Ricci} the following gauge equivalence\footnote{A version of this description was proposed in~\cite{Bekaert:2013zya,Alkalaev:2014nsa}.}
\begin{align}\label{gaugeamb}
    G_{AB}\sim G_{AB}+\lambda \,G_{AB} +\lambda_{(A} V_{B)} +\lambda_{AB}\, V^2\,,
\end{align}
where $\lambda$, $\lambda_A$ and $\lambda_{AB}$ are gauge parameters. The parameter $\lambda$ is related to the usual Weyl symmetry  while $\lambda_A$ and $\lambda_{AB}$ implement the equivalence relation up to components along the homothety vector field and up to terms vanishing on the null-cone $V^2=0$.\footnote{Note that this formulation (more precisely, its parent version) of the off-shell conformal gravity  gives a manifestly $\mathfrak{so}(d,2)$-covariant description of the respective jet-space and BRST complex employed~\cite{Boulanger:2004eh,Boulanger:2004zf} in classifying conformal invariants.}  Both  interpretations of the system \eqref{sp2eqA}-\eqref{Ricci} as well the system \eqref{sp2eqA}, \eqref{gaugeamb} have simple toy model counterparts in the context of the scalar field in ambient space, described in Appendix \bref{sec:off/on}.

The condition \eqref{Ricci} of Ricci flatness on the ambient metric in the FG construction can be understood as a gauge-fixing condition for the arbitrariness \eqref{gaugeamb} in the extension of the metric from the projective null cone to the whole ambient space. This gives a field-theoretic explanation of the relation between the two equivalent ambient descriptions of off-shell conformal gravity.

One can even try to exploit the relation even further and to interpret the ambient system \eqref{sp2eqA}, \eqref{gaugeamb} as  defining a field theory on the curved hyperboloid $V^2=-1$. We postpone detailed discussion of this approach till section~\bref{sec:factorized-system} and only mention that there is a simple example illustrating this idea. Consider the following ambient system:
\begin{equation}
(V^A\d_A+\Delta)\Phi=0\,, \qquad \Phi\sim \Phi+V^2\lambda\,,
\end{equation}
where $\Phi$ is a scalar field on ambient space. Interpret this system as defining a scalar field on the curved hyperboloid $V^2=-1$ rather than on the projective hypercone $V^2=0$. In so doing one can try to assume $\Phi$ harmonic  by adding terms proportional to $V^2$, i.e. try to pick a representative which obeys $\nabla^2\Phi=0$. Of course, this is a subtle procedure and for it to work properly one needs to be careful with functional issues.

To complete the discussion of the gauge equivalence~\eqref{gaugeamb} let us note that~\eqref{gaugeamb} can be compactly written in terms of the generating functions $F_i$\,:
\begin{equation}
\label{equiv-F}
F_+\sim F_++ \lambda \,F_++\lambda^A P_A\,\,F_0+\frac12\,\lambda^{AB}P_AP_B\,\,F_-\,.
\end{equation}
In this form it is clear that this equivalence corresponds to the usual equivalence of constrained systems related to an infinitesimal redefinition of the constraints. This system (more precisely, its BRST extension) was considered in~\cite{Grigoriev:2006tt,Alkalaev:2014nsa} in the context of AdS HS gauge theory.

%%%%%%%%%%%%%%%%%%%%%%%%%%%%%%%%%%%%%%%%%%%%%%%%%%%%%%%%%%%%%%%%
\section{Higher-spin Extension of Fefferman-Graham Theory}
\label{sec:hsextension}
%%%%%%%%%%%%%%%%%%%%%%%%%%%%%%%%%%%%%%%%%%%%%%%%%%%%%%%%%%%%%%%%

The idea to describe HS fields by allowing all powers of momenta in the $\mathfrak{sp}(2)$ constraints was at the core of Bars' proposal in \cite{Bars:2001um}. Accordingly, a bold guess
 for a HS extension of the FG construction is to remove the restriction on $F_i$ to contain no more than two powers of ambient momentum $P_A$:
\begin{align}
\label{HSext}
    \begin{aligned}
        F_+(X,P)&=\Phi(X)+\Phi^A(X) P_A+\half {G^{AB}}(X) P_A P_B+\Phi^{ABC}(X) P_A P_B P_C+\ldots\,,\\
        F_0(X,P)&=V(X)+{V^A}(X)P_A+\half {V^{AB}}(X) P_A P_B+\ldots \,,\\
        F_-(X,P)&={\tilde G}(X)+{\tilde G}^A(X)P_A+\ldots\,,
    \end{aligned}
\end{align}
where the dots denote possible terms of higher order in the momenta. Moreover, one should require $G^{AB}(X)$ and $V^A(X)$ to remain, respectively, nondegenerate and nowhere vanishing. This requirement, or the choice of a particular vacuum, will break the apparent democracy between the three constraints $F_i(X,P)$ even if one allows for arbitrary dependence on the momentum for all the three constraints.

We assume that the equations of motion remain the same, which makes the system consistent with an arbitrary gravitational background (i.e. $G_{AB}$ and $V^C$): 
\begin{equation}
\label{sp2eq}
 \pb{F_i}{F_j}=C^k_{ij}F_k\,.
\end{equation} 
Here $C^k_{ij}$ are $\mathfrak{sp}(2)$ structure constants. The gauge symmetries are
\begin{equation}
\label{sp2gs}
\delta_\epsilon F_i=\pb{\epsilon}{F_i}\,.
\end{equation} 

In what follows we analyze this system and relate it to other formulations of HS gauge fields. As we are going to show, the same system with the Poisson bracket replaced with the Weyl star-commutator is more appropriate in the context of HS fields. Although the above system has something to do with both conformal HS fields in $d$ dimensions and HS fields in $d+1$ dimensions, here we are mostly interested in the ($d+1$)-dimensional interpretation.

The full system~\eqref{sp2eq}, \eqref{sp2gs} with $F_i$ as in~\eqref{HSext} is known in the literature  in various versions (see e.g.~\cite{Bars:2001um,Grigoriev:2006tt}). In the context of HS fields in $d+1$ dimensions, precisely this system was proposed in~\cite{Grigoriev:2012xg} where it was proved to describe off-shell HS fields upon linearizing over the AdS background solution. Note also that a version of this system with two generators fixed and $F_+$ involving higher powers in $P$ was also suggested in \cite{Bonezzi:2010jr} to describe HS fields in $d$ dimensions  and, moreover, the $\mathfrak{sp}(2)$ system was cast into an  action principle in \cite{Bonezzi:2014nua}.

The above construction can also be phrased in terms of tractor fields which can be identified with certain tensor fields in ambient space restricted to the hypercone 
\cite{Cap:2002aj,Gover:2004as, Gover:2006qm}.  In particular, this language was employed to described HS fields in  \cite{Gover:2008sw,Gover:2008pt,Gover:2009vc,Gover:2009,Grigoriev:2011gp}.

%%%%%%%%%%%%%%%%%%%%%%%%%%%%%%%%%%%%%%%%%%%%%%%%%%%%%%%%%%%%%%%%
\subsection{Off-shell Higher-spin Fields on  Gravitational Backgrounds}
\label{sec:offgra}
%%%%%%%%%%%%%%%%%%%%%%%%%%%%%%%%%%%%%%%%%%%%%%%%%%%%%%%%%%%%%%%%

Suppose we are given a triplet of ambient functions $F^0_+(X,P)$, $F^0_0(X,P)$, $F^0_-(X,P)$ that are of degree
$2,1,0$ in $P_A$, respectively, and whose Poisson brackets satisfy  the algebra $\mathfrak{sp}(2)$. We assume that $F^0_+(X,P)$ defines a nondegenerate  metric and $F^0_0(X,P)$ defines a nowhere-vanishing vector field. Then, as was mentioned before, it follows that they are of the form~\eqref{sp2constr} with the scalar field $F(X)$ given by \eqref{sp23} and the vector field $V^A(X)$ satisfying $\nabla_A V^B=\delta^A_B$.

As was shown by Fefferman and Graham, at least locally any metric $g_{\mu\nu}$ in $d+1$ dimensions can be lifted to an ambient metric
$G_{AB}$ defined on the ($d+2$)-dimensional ambient space such that the original spacetime is a ``curved hyperboloid'' determined by $G_{AB}V^AV^B=-1$,
while the original metric $g_{\mu\nu}$ is a pullback of $G_{AB}$ to this surface.

We are going to interpret $F_i=F^0_i$ as a background solution of the full HS system, where only gravitational fields are nonvanishing.
In order to see that the full system indeed describes HS fields, let us consider its linearization around $F_i=F^0_i$. The linearized equations and gauge symmetries read:
\begin{equation}\label{deltafi}
 \pb{f_i}{F^0_j}+\pb{F^0_i}{f_j}=C_{ij}^k f_k\,, \qquad \delta_\epsilon f_i=\pb{\epsilon}{F^0_i},
\end{equation} 
where $f_i$ is a perturbation. In more detail, for the gauge transformations of $f_0$ and $f_-$ one gets
\begin{align}
    \delta_\epsilon f_0&=\pb{\epsilon}{F^0_0}=(V \cdot D-P\cdot \d_P) \epsilon\,,& 
    \delta_\epsilon f_-&=\pb{\epsilon}{F^0_-}=({V}\cdot \d_P) \epsilon\,,
\end{align}
where $D_M$ denotes the covariant derivative acting on generating functions: $D_M=\dl{X^M}+\Gamma_{MN}^KP_K\dl{P_N}$ (see also Appendix \bref{covder}).
Since $V^A$ is nowhere vanishing, this implies that the gauge $f_-=f_0=0$ is reachable. Indeed,
by picking a suitable coordinate system one can always assume that the homothety vector field reads $V^A\frac{\partial}{\partial X^A}=\dl{\rho}$, where $\rho$ is one of the coordinates. In such a coordinate system it is clear that using suitable $\epsilon$ one can set both $f_{-}$ and $f_0$ to zero. 

Although the above  argument applies to the linearized system, it extends in a usual way to the nonlinear level  provided one restricts oneself to solutions which are ``sufficiently close" to $F^0_i$. 
For such solutions, the two remaining $\mathfrak{sp}(2)$ relations involving the perturbation $f_+$ (not necessarily small) imply:
\begin{equation}
    (P\cdot \d_P-V \cdot D-2)f_+=0\,,\qquad 
    ({V}\cdot \d_P) f_+=0\,,
\end{equation}
where $F_+=F^0_++f_+$. While the first equation uniquely determines $f_+$ in terms of its value on the level surface $V^2=\mathrm{constant}$, the second one implies that $f_+$ does not depend on the components of $P$ along $V$. Finally, such $f_+$ are in one-to-one correspondence with totally symmetric tensor fields on the level surface $V^2=\mathrm{constant}$.

Similarly, in the gauge $F_0=F^0_0$ and $F_-=F^0_-$ the residual gauge symmetries read as
$$
\begin{gathered}\label{deltaf2}
{\delta_\epsilon f_+=\pb{\epsilon}{F_+}=(P\cdot D) \epsilon} +\pb{\epsilon}{f_+}\,,  \quad\mbox{with}\quad  (V \cdot \d_P) \epsilon=0\,, \quad (P\cdot \d_P-{V}\cdot D)\epsilon=0\,.
\end{gathered}
$$
The global reducibility parameters $\epsilon_0$ for the linearized symmetries in \eqref{deltafi} are described by the $\mathfrak{sp}(2)$ centralizer equations $\pb{\epsilon_0}{F_i^0}=0$ which read more explicitly: 
\begin{equation}\label{offHS}
(P\cdot D) \epsilon_0=(P\cdot \d_P-{V}\cdot D)\epsilon_0=(V \cdot\d_P)\epsilon_0=0\,.
\end{equation}
These global symmetries (= global reducibility parameters) define an ``off-shell HS algebra'', by which we simply mean the $\mathfrak{sp}(2)$ centralizer \eqref{offHS} above, canonically equipped with a Lie algebra structure via the Poisson bracket. More invariantly, the off-shell HS algebra arises here in a usual way (see e.g.~\cite{Barnich:2001jy} for more details) as an algebra of gauge transformations preserving a given vacuum solution. Its Lie bracket comes from the commutator of the respective gauge transformations.
In particular, for the flat vacuum $V^A=X^A$ we get a version of the standard off-shell HS algebra.

We comment more on various off-shell HS algebras below when the Poisson bracket is replaced by the star-product.
 It is worth stressing that the off-shell HS algebra is defined on any gravitational background, but it is generically trivial.
In fact, the system of equations \eqref{offHS} describes totally symmetric Killing tensors in $d+1$ dimensions, which may not admit nontrivial solutions on a generic background.

To summarize, the natural HS extension of the FG off-shell system provides an elegant description of totally symmetric tensor gauge fields in $d+1$-dimension subject to nonlinear gauge symmetries in an arbitrary gravitational background.\footnote{This does not contradict the usual no-go theorems (such as \cite{Aragone:1979hx}) on the propagation of HS gauge fields in generic gravitational backgrounds since here the tensor gauge fields are \textit{off-shell}. For instance, at linearized level the covariantization of Fronsdal gauge transformations is perfectly well defined. It is only its compatibility with the field equations which is problematic.}

%%%%%%%%%%%%%%%%%%%%%%%%%%%%%%%%%%%%%%%%%%%%%%%%%%%%%%%%%%%%%%%%
\subsection{Poisson Bracket vs. Star-Product}
\label{sec:poisstar}
%%%%%%%%%%%%%%%%%%%%%%%%%%%%%%%%%%%%%%%%%%%%%%%%%%%%%%%%%%%%%%%%

If we are interested in the
\textit{on-shell} fields, then the construction of the previous section is not entirely satisfactory. Moreover, even at the off-shell level the corresponding algebra of global symmetries coincides with the familiar off-shell HS algebra as a linear space, but is a Lie algebra with respect to the Poisson bracket, rather than the Weyl commutator as it should.
This suggests that the construction has to be modified by 
replacing the Poisson bracket with the Weyl $\star$-commutator determined by 
\begin{equation}
    \qcommut{X^A}{P_B}=\delta^A_B\,.
\end{equation}

Another reason for considering the ``quantum'' version has to do with the interpretation of the off-shell nonlinear system as describing  background conformal HS fields in $d$ dimensions to which a scalar field can consistently couple (for more details, see e.g. \cite{Segal:2002gd,Tseytlin:2002gz,Bekaert:2010ky,Bekaert:2013zya,Grigoriev:2016bzl,Bonezzi:2017mwr}). Note that from this perspective the Poisson bracket version naturally corresponds to background fields for the point particle. This in turn is a mechanical model whose wave function is the above scalar field. The $\star$-product is crucial for the scalar field to couple to arbitrary HS background fields. It arises directly as a quantization of the phase space $\{x^a,p_b\}=\delta^a_b$ where HS background fields correspond to arbitrary functions $f(x,p)$. The $\mathfrak{sp}(2)$-system provides an ambient space extension of this construction \cite{Grigoriev:2006tt,Bekaert:2009fg,Bekaert:2013zya}.

Therefore, we pass to the $\star$-commutator instead of the Poisson bracket and the corresponding version of the basic system \eqref{sp2eq} reads as
\begin{equation}
\label{sp2eq-star}
 \qcommut{F_i}{F_j}=C^k_{ij}F_k\,, \qquad 
\delta_\epsilon F_i=\qcommut{\epsilon}{F_i}\,.
\end{equation} 
Note that if one restricts to the spin-two version of this $\mathfrak{sp}(2)$-system, then the same solution \eqref{sp2constr}, \eqref{sp23}, \eqref{FGsp2} that satisfies the Poisson bracket version \eqref{sp2eqA} also solves \eqref{sp2eq-star}. This system of operators plays an important role in the ambient description of scalar fields~\cite{GJMS} (see also~~\cite{Gover:2002ay,Manvelyan:2007tk}), in particular for the singleton (see Appendix \bref{sec:off/on}).

Let us now restrict ourselves to the flat vacuum solution: $G^{AB}=\eta^{AB}$, $V^A=X^A$. It is easy to see that in this case $F^0_i$ solves the above equations and hence gives a vacuum solution. Linearizing the equations and the gauge symmetries around $F^0_i$ one gets
\begin{align}
\qcommut{F^0_i}{f_j}+\qcommut{f_i}{F^0_j}&=C_{ij}^k f_k & \delta_\epsilon f_i&=\qcommut{\epsilon}{ F^0_i}\,.
\end{align}
It follows again that $f_{0}$ and $f_-$ can be gauged away,
resulting in the linearized system\footnote{Note that had we linearized the system around general gravitational background the linearized gauge transformations for a spin $s$-field would in general not only involve contributions from parameters of spin $s-1$ but also those with lower spin. This
mixing can be traced back to nonlinear in $X$ terms in $F^0_+$ involved in the star-commutator with the parameter. It is similar to the analogous mixing observed~\cite{Grigoriev:2016bzl} in the case of conformal HS fields on the boundary and is consistent with the fact that CHS fields are boundary values of the bulk ones. We are grateful to A.~Tseytlin for discussion of this point.}
\begin{equation}\label{offsys}
\begin{gathered}
(X\cdot \d_P) f_+=0\,, \qquad (P\cdot \d_P-{X}\cdot \d_X-{2})f_+=0\,,
\\
{\delta_\epsilon f_+=\qcommut{\epsilon}{F^0_+}=(P\cdot \d_X) \epsilon}\,, \qquad  (X\cdot \d_P) \epsilon=0\,, \quad (P\cdot \d_P-{X}\cdot \d_X)\epsilon=0\,.
\end{gathered}
\end{equation}

To see the relation with Fronsdal fields, let us recall their ambient space formulation. In terms of the generating function
$\Phi(X,P)=\sum_s\frac{1}{s!}\,\Phi^{A_1\ldots A_s}(X)\,P_{A_1}\ldots P_{A_s}$\,,
the equations of motion and gauge symmetries read as
\begin{eqnarray}
\label{Fronsdal-ambient}
{(\d_X\cdot \d_X) \Phi=(\d_{P}\cdot\d_{X})\Phi=(\d_{P}\cdot\d_{P})\Phi=0}\,,\qquad  {\delta_\epsilon\Phi=(P \cdot \d_X) \epsilon}\,,\\
(X\cdot \d_X-P\cdot\d_P+2)\Phi={(X \cdot\d_{P})\Phi}=0\,,\label{Fr-amb}
\end{eqnarray}
which is equivalent to 
\begin{align}\label{Fronsdal_d+1}
(\nabla^2-m_s^2)\phi_{\mu_1\ldots \mu_s}=\nabla^\mu\phi_{\mu\mu_2\ldots \mu_s}=\phi^\mu{}_{\mu \mu_3\ldots\mu_s}=0\,,\qquad \delta_\epsilon \phi_{\mu_1\ldots \mu_s}&=\nabla_{(\mu_1}\epsilon_{\mu_2\ldots\mu_s)}\,,
\end{align}
in terms of the fields defined on the hyperboloid, where $m_s$ is the mass of spin-$s$ Fronsdal field on $(A)dS_{d+1}$.

It is clear from \eqref{Fronsdal-ambient} that, in order to put the linearized off-shell system \eqref{offsys} on-shell, one has to impose
the equation of motion $(\d_X\cdot\d_X) f_+=0$ (the remaining divergence and trace constraints arise automatically as consistency conditions). The above analysis of the system~\eqref{sp2eq-star} and its relation to Fronsdal fields has been originally performed in~\cite{Grigoriev:2012xg} using the parent formulation technique.

To draw an analogy with the FG description of gravity note that $(\d_X\cdot\d_X) f_+=0$ is a linearized HS analogue of the Ricci flatness $R_{AB}=0$ condition. To find a nonlinear HS version of the Ricci flatness remains a tantalizing open problem in the metric-like formulation. We pursue a different approach in the next section.

%%%%%%%%%%%%%%%%%%%%%%%%%%%%%%%%%%%%%%%%%%%%%%%%%%%%%%%%%%%%%%%%
\section{Towards On-shell Higher-Spin Theory}
\label{sec:hsonshell}
%%%%%%%%%%%%%%%%%%%%%%%%%%%%%%%%%%%%%%%%%%%%%%%%%%%%%%%%%%%%%%%%
The $\mathfrak{sp}(2)$-system 
captures off-shell backgrounds, both gravitational and HS ones. In the case of gravity, the ($d+1$)-dimensional Einstein equations with cosmological constant result from Ricci flatness in the ($d+2$)-dimensional ambient space. A natural question is whether it is possible to directly put the off-shell HS system~\eqref{sp2eq-star} on-shell.

One possible way would be to find nonlinear corrections
to the constraints in the first line of \eqref{Fronsdal-ambient}. However, it is not clear which structures may control such deformation and in any case in this way there is no obvious way to reconstruct the system nonperturbatively.

An alternative is to further exploit the analogy with constrained systems. The $\mathfrak{sp}(2)$ relations imposed on $F_i$ can be interpreted as a condition that $F_i$
are first-class constraints while gauge transformations $\delta_\epsilon F_i=\qcommut{\epsilon}{F_i}$ correspond to canonical transformations. The general first-class condition (= closure of the algebra)
\begin{equation}\label{ext1}
[F_i,F_j]_\star=U^k_{ij} \star F_k
\end{equation}
is preserved by the following
gauge transformations
\begin{equation}\label{ext2}
\delta F_i=\lambda^j_i \star F_j\,,
\end{equation}
in addition to $\delta_\epsilon F_i=\qcommut{\epsilon}{F_i}$.
The transformations \eqref{ext2} correspond to infinitesimal redefinitions of the constraints (at classical level such symmetries preserve the constraint surface). The $\mathfrak{sp}(2)$ system \eqref{sp2eq-star} can be seen as a partial gauge-fixing of the system \eqref{ext1}-\eqref{ext2}.\footnote{Strictly speaking one needs to prove that such gauge is reachable. Although this is easy to see at the linearized level and hence this is also true for the field configurations that are sufficiently close to the vacuum, a general statement is not known.} 

To explain what symmetries~\eqref{ext2} can be useful for, let us consider the linearization of the system about the flat vacuum solution $F^0_i$ corresponding to $G^{AB}=\eta^{AB}$ and $V^A=X^A$,
i.e. $F^0_+=\frac12P\cdot P$, $F^0_0=X\cdot P$ and $F^0_-=-\frac12 X\cdot X$. 
The linearization of 
\eqref{ext2} reads as
\begin{equation}
\delta f_i=\lambda^j_i\star F^0_j\,.
\end{equation}
At the formal level, these symmetries can be employed to make $f_i$ satisfy the constraints in the first line of \eqref{Fronsdal-ambient}, i.e. $(\d_X\cdot \d_X) f_i=(\d_{P}\cdot\d_{X})f_i=(\d_{P}\cdot\d_{P})f_i=0$. For instance, in the space of polynomials in $X,P$ variables this is exactly the case.

As we have seen $f_0$ and $f_-$ can be set to zero by the gauge symmetries $\delta f_i=\qcommut{\epsilon}{F^0_i}$ (for $i=0\,,-$). As a result, there is only one field left, $f_+$. It satisfies  \eqref{Fronsdal-ambient}-\eqref{Fr-amb} and hence describes the Fronsdal system \eqref{Fronsdal_d+1}.
The problem with this formal argument is that the space of polynomials in $X,P$ is not the one where one can set to zero $f_{0}$, $f_-$ (since one somehow has to ``invert'' the operators  $X\cdot\partial_P$ and $X\cdot \pl_X-P\cdot \pl_P$). Moreover,
it is not the functional space relevant for describing field theory configurations.
Other way around, the space (natural from the field theory perspective) of polynomials in $P$ and smooth functions in $X$ defined in the vicinity of $X^2=-1$ allows one to eliminate $f_{0}$, $f_-$ but, in this space, the gauge transformations \eqref{ext2} can be used to set to zero all the fields $f_i$.

It turns out that one can nevertheless use this system to describe on-shell fields by reformulating the system in parent form (see  section \bref{parentref}) and requiring the fields to belong to a special functional class.

A heuristic explanation for why the extended system \eqref{ext1}-\eqref{ext2} is capable of describing on-shell HS gauge fields in the bulk employs boundary analysis. More precisely, the extended system in the vicinity of the hypercone $V^2=0$ is known \cite{Bekaert:2013zya,Alkalaev:2014nsa} to describe off-shell conformal HS gauge fields in $d$ dimensions. In their turn, these off-shell conformal fields are boundary values of the on-shell bulk fields in $d+1$ dimensions. However, in the ambient approach bulk fields and their boundary values are described by exactly the same ambient system, considered either near $V^2=-1$ or near $V^2=0$ (for more details see~\cite{Bekaert:2012vt,Bekaert:2013zya}). This justifies that the above extended system has something to do with on-shell bulk fields provided one considers it near $V^2=-1$.

To give the above considerations a precise meaning it is useful to reformulate the system using the parent approach which has proved to be instrumental in analyzing boundary values~\cite{Bekaert:2012vt,Bekaert:2013zya,Chekmenev:2015kzf}.

%%%%%%%%%%%%%%%%%%%%%%%%%%%%%%%%%%%%%%%%%%%%%%%%%%%%%%%%%%%%%%%%
\subsection{Parent Reformulation}\label{parentref}
%%%%%%%%%%%%%%%%%%%%%%%%%%%%%%%%%%%%%%%%%%%%%%%%%%%%%%%%%%%%%%%%
The system~\eqref{sp2eq-star} can be equivalently reformulated in  parent form. The underlying idea is to put the constrained system under consideration in an auxiliary space where genuine space-time coordinates are replaced by formal variables, typically denoted by $Y^A$, so that the equations of motion and gauge symmetries become purely algebraic.  The parent formulation is then constructed as a gauge field theory whose target space is the space of fields on the auxiliary space while the source space is the original space-time. Moreover, in so doing the gauge parameters are promoted to 1-form fields of the parent system while reducibility parameters to $p$-form fields. In other words, parent formulation is an AKSZ sigma model~\cite{Alexandrov:1995kv} whose target space is the jet-space BRST complex of the intial theory.  The equivalence with the original formulation is maintained by imposing free differential algebra relations and their associated gauge symmetries, on top of the auxiliary space version of the original equations of motion and gauge symmetries. More details and references can be found in \cite{Barnich:2004cr,Barnich:2006pc,Barnich:2010sw,Grigoriev:2010ic}.

Being a sigma model of AKSZ type (or equivalently  a free differential algebra with constraints) the parent formulation can be consistently pulled back to any space-time submanifold or even put (at least formally) on a totally different space-time manifold. This property is extremely useful in analyzing the relation between the Hamiltonian and Lagrangian description~\cite{Grigoriev:1999qz}, manifest realization of symmetries~\cite{Barnich:2006pc,Bekaert:2009fg} boundary values of gauge fields~\cite{Vasiliev:2012vf,Bekaert:2012vt}, and, more generally, holographic dualities~\cite{Vasiliev:2012vf,Bekaert:2013zya}.

In the case at hand the parent reformulation is constructed by first extending $X^A$ and $P_B$ with additional variables $Y^A$.\footnote{ The geometric meaning is that $Y^A$ are coordinates on the fibres of the tangent bundle over the ambient space.} Now $X^A$ are the usual space-time coordinate, while $Y^A$ and $P_B$ are auxiliary variables needed to conveniently pack fields into generating functions. Note however, that Weyl $\star$-product is now in the space of $Y,P$ variables so that spacetime coordinates are not explicitly involved. The field content consists of the original constraints $F_i(X,Y,P)$ and a new field, the connection one-form $A=dX^B A_B(X,Y,P)$ associated to the original gauge parameter. The fields $F_i$, $A$ are interpreted as generating functions for the component fields identified as the respective coefficients in the expansion over $Y,P$.

Due to the fact that the parent formulation contains an infinite number of fields the specification of the functional class of $F_i$ and $A$ in the $Y$ space is part of the definition of the system. The minimal choice to begin with  is that of formal power series in $Y$. With this choice the parent formulation is equivalent to \eqref{sp2eq-star}. Let us for definiteness and simplicity also restrict ourselves to polynomials in $P$, so that we are dealing with polynomials in $P$ with coefficients in formal power series in $Y$.  

The parent form of the $\mathfrak{sp}(2)$ system, the equations of motion we are going to study, read~\cite{Grigoriev:2012xg}
\begin{equation}
\label{parent-d+2}
\begin{aligned}
 dA-\half\qcommut{A}{A}&=0\,,  &dF-\qcommut{A}{F}&=0\,, & \qcommut{F_i}{F_j}-C^k_{ij}F_k&=0\,,\\
 \delta_\epsilon A&=d\epsilon-\qcommut{A}{\epsilon}\,, & \delta_\epsilon F_i&=\qcommut{\epsilon}{F_i}\,,
 \end{aligned}
\end{equation} 
where from now on $\qcommut{\cdot}{\cdot}$ denotes the Weyl $\star$-commutator (in the $Y,P$-space). The classical limit where the star-product commutator is replaced by the Poisson bracket in the $Y,P$-space also makes sense but, as we argued above, its interpretation from the effective action point of view is unclear.

The above parent system \eqref{parent-d+2} is background independent and can be considered on ($d+1$)-dimensional manifolds. 
This can be done by pulling-back the system~\eqref{parent-d+2} to the curved hyperboloid described by $V^2=-1$. The advantage of the parent formulation is that $V^A$ is non-dynamical and can be conveniently manipulated. In particular, in the resulting theory defined on $V^2=-1$ one can gauge fix $V^A$ such that $V^A$ is constant.  Furthermore, in contrast to the $X,P$-space of the previous sections, the auxiliary $Y,P$-space is flat and it is easier to impose algebraic conditions in order to put the system on-shell. 

There is a parent realization for the anti-de Sitter solution of the HS extension of the FG construction, which reads:~\cite{Grigoriev:2006tt,Grigoriev:2012xg}
\begin{eqnarray}
\label{standard-F-vacuum}
%\begin{gathered}
 &F^0_+=\half P\cdot P\,, \qquad F^0_0=\frac12\{(V+Y)^A, P_A\}_\star\,, \qquad F^0_-= \half (V+Y)\cdot (V+Y)\,,& \\
 &A^0=\omega^A_B\,(V^B+Y^B)P_A\,,&
% \end{gathered}
\label{standard-F-vacuum2}
\end{eqnarray} 
where $V^A$ is constant, $\{\,\,,\,\}_\star$ denotes the $\star$-anticommutator and $\omega$ is a standard $\mathfrak{so}(d,2)$ flat connection such that $\nabla_\mu V^B$ has rank $d+1$. Note that there are more general solutions where, rather than \eqref{standard-F-vacuum2}, $A_0$ is taken to be any flat connection taking values in the off-shell HS algebra, i.e. $dA_0-\half\qcommut{A_0}{A_0}=0$ and $[F^0_i,A^0]_\star=0$, so that $A^0$ is polynomial in $Y$.  Note that although it is easy to check that  
\eqref{standard-F-vacuum}-\eqref{standard-F-vacuum2} is a solution by redefining variables $Y^A+V^A \to Y^A$, this redefinition is not well-defined for formal power series in $Y$. In particular, the shift by $V$ crucially affects the content of the theory.

The gauge symmetries leaving the vacuum solution \eqref{standard-F-vacuum}--\eqref{standard-F-vacuum2} intact,
\begin{align}
    d\epsilon^0-[A^0,\epsilon^0]_\star&=0\,, & [F^0_i,\epsilon^0]_\star&=0\,,
\end{align}
are $1:1$ with the off-shell HS algebra defined as the algebra of elements $\star$-commuting with all $F^0_i$'s. In contrast to section \bref{sec:offgra}, this off-shell HS algebra is not just a Lie algebra, but  is an associative one because the $\star$-product of two $\mathfrak{sp}(2)$-singlets is a singlet again. This off-shell algebra is directly related to the symmetries of the conformal Laplacian described by Eastwood \cite{Eastwood:2002su}, i.e. it has a two-sided ideal that can be quotiented out as to get the on-shell HS algebra.

Let us consider the linearization of the parent system around \eqref{standard-F-vacuum}. The linearized fluctuations may be required to be totally traceless in which case one finds an on-shell version. This is analogous to imposing $(\pl_X\cdot \pl_X) f_+=0$ in section \bref{sec:poisstar}. More precisely, it can be shown \cite{Grigoriev:2012xg} that requiring the linearized fluctuations of \eqref{parent-d+2} to be in the kernel of
$\pl_Y\cdot \pl_Y$, $\pl_Y\cdot \pl_P$, and $\pl_P\cdot \pl_P$ results in the free Fronsdal equations.  The problem is to extend this beyond the linearized approximation. For spin-two, the Ricci flatness provides a nonlinear completion of these equations but its HS analogue is still missing.

%%%%%%%%%%%%%%%%%%%%%%%%%%%%%%%%%%%%%%%%%%%%%%%%%%%%%%%%%%%%%%%%
\subsection{Factorization}
\label{sec:factorized-system}
%%%%%%%%%%%%%%%%%%%%%%%%%%%%%%%%%%%%%%%%%%%%%%%%%%%%%%%%%%%%%%%%
We can push the on-shell HS extension of the FG construction a bit further and get equations that describe propagation of HS fields on any HS-flat background. Indeed, there is a formally consistent factorization of the system by the ideal generated by $F_i$. The factorization is obtained by imposing the extra gauge symmetry:
\begin{equation}\label{tracesym}
 \delta A=\lambda^i\star F_i\,,\qquad\qquad \delta F_i= \lambda^j_i\star F_j\,,\qquad\qquad 
\end{equation}
where $\lambda^i,\lambda^i_j$ are gauge parameters, and requiring equations of motion to hold modulo terms  proportional to $F_i$. The full system of equations that
is suitable for describing on-shell fields together with gauge transformations is~\cite{Alkalaev:2014nsa} (see also~\cite{Grigoriev:2006tt} for earlier versions)
\besubeqs
\label{full-parent}
\begin{align}
 dA-\half\qcommut{A}{A}&= u^i\star F_i\,, &
 \delta A&=d\epsilon-\qcommut{A}{\epsilon}+\lambda^j\star F_j \,,\\
 dF_i-[A,F_i]_\star&= u^j_i\star F_j\,, &\delta F_i&=\qcommut{\epsilon}{F_i}+ \lambda^j_i \star F_j\,,\\ \qcommut{F_i}{F_j}-C_{ij}^kF_k&= u_{ij}^k \star F_k\,.
\end{align} 
\esubeqs
Here $u$'s are non-dynamical fields that transform under $\epsilon$, $\lambda^j_i$, $\lambda^j$ in an obvious way. Note that $u^i_{jk}$ are not unconstrained and have to obey the relations following from the Jacobi identities. {This system is a candidate for the on-shell HS-extended FG theory.}

Now we are going to study the above system perturbatively over a HS-flat vacuum solution, where $F_i=F^0_i$ as in~\eqref{standard-F-vacuum} while $A^0$ is more general. To this end we introduce an appropriate functional class that allows one to have gauge symmetry \eqref{tracesym} without trivializing the solution space and such that the off-shell HS algebra gets reduced to the correct on-shell HS algebra. The functional class  $\mathfrak{C}$ is that of polynomials in $P$ with coefficients that are formal power series in $Y$.\footnote{It is important to distinguish $\mathfrak{C}$ from the space of formal power series in $Y$ with coefficients that are polynomials in $P$.} Having formal power series in $Y$ is important for being able to gauge away fluctuations of $F_{-}$, $F_0$. We also require that $\mathfrak{C}$ is of finite trace order, i.e. for any $f(Y,P)\in \mathfrak{C}$ there exists an $\ell\in\mathbb N$ such that 
\begin{align}
{(\d_Y\cdot \d_Y)^\ell f=0\,.}\label{ftr}
\end{align}
Note that the space of functions of finite trace order is a module over polynomials in $Y,P$, i.e. we can $\star$-multiply $f(Y,P)\in \mathfrak{C}$ by a polynomial $p(Y,P)$ and the result, $f(Y,P)\star p(Y,P)$, is still in $\mathfrak{C}$. It then follows that any function in $\mathfrak{C}$ can be decomposed as
\begin{align}
f=f_0+f^i_{1}\star F^0_i + f^{ij}_{2} \star F^0_i\star F^0_j+\ldots\,, \qquad\qquad f_{n}\text{ -- totally traceless}
\,,
\end{align}
such that the number of terms is finite. Having an element $f$ decomposed as above, we define a projector onto the traceless part:\footnote{Note that the trace decomposition is defined with respect to $F^0_-=-\tfrac12(Y+V)\cdot (Y+V)$, $F^0_0=P\cdot (Y+V)$, and $F^0_+=\tfrac12P\cdot P$ i.e. this is not the usual decomposition into traceless tensors $f=f_0+ Y\cdot Y f_1+...$ due to the shift by $V$ and due to the $\star$-product.}
\begin{align}
{\Pi f=f_0}\,.
\end{align}

Now we can linearize the system and put it on-shell.  As anticipated above, we take a slightly more general vacuum, where $A^0$ does not have to be just the flat connection \eqref{standard-F-vacuum2} linear in $P$, describing $AdS_{d+1}$.  The nontrivial part of the vacuum equations reads
\besubeqs
\begin{align}\label{vaca}
 dA^0-\half\qcommut{A^0}{A^0}= u^i\star F^0_i\,,\\
 [A^0,F_i^0]_\star= u^j_i\star F^0_j\,.\label{vacb}
\end{align}
\esubeqs
It follows that $A^0$ is equivalent to a flat connection of the on-shell HS algebra. Indeed, the on-shell HS algebra of Eastwood \cite{Eastwood:2002su} can be defined  within the present framework as follows:\footnote{An oscillator realization of this algebra was given in \cite{Vasiliev:2003ev}.} 
\begin{align}\label{onshellHSalg}
\chi\in \text{on-shell HS-algebra}&: &[\chi,F^0_i]_\star&=0\,, &&\chi\sim \chi+\lambda^i \star  F^0_i \,,
\end{align}
where $\chi$, $\lambda^i$ are in $\mathfrak{C}$. The $\mathfrak{sp}(2)$-singlet constraints solved for $\chi$ in $\mathfrak{C}$ imply that $\chi$ is a polynomial in $Y$.

Next, for $A^0$ entering~\eqref{vaca}-\eqref{vacb}, we can use  gauge symmetry \eqref{tracesym} on the vacuum solution
\begin{equation}\label{tracesym2}
 \delta A^0= \lambda^i\star F^0_i \,,
\end{equation}
as to gauge away all traces and arrive at $A^0(Y,P)$ satisfying $\d_Y\cdot \d_Y A^0=\d_Y\cdot \d_P A^0=\d_P\cdot \d_P A^0=0$ (i.e. $A^0$ is a collection of traceless tensors in $Y$ and $P$, or equivalently $\Pi A^0=A^0$). Then the traceless part of \eqref{vaca}-\eqref{vacb} implies that $A^0$ is a flat connection of the on-shell HS algebra:
\begin{align}\label{Flaths}
    dA^0&=\half\Pi\big(\qcommut{A^0}{A^0}\big)\,.
\end{align}
The gauge symmetries preserving the vacuum solution are determined by 
\begin{align}
    d\epsilon^0&=[A^0,\epsilon^0]_\star+ \lambda^i\star F^0_i\,, & [F^0_i,\epsilon^0]_\star+\lambda^j_i\star F^0_j &=0\,.
\end{align}
We can again decompose $\epsilon^0$ into the trace part that is proportional to $ F^0_i$ and the traceless part. The trace part unambiguously fixes the $\lambda$'s, while the traceless part is covariantly constant with respect to $A^0$. Therefore, the global symmetry parameters are parameterized by the on-shell HS algebra, as it should be.

Now we can study fluctuations over $A^0$ and $F^0_i$ and determine the general structure of the equations. Let us expand
\begin{align}
A=A^0+a\,, \qquad F_i=F^0_i+f_i\,,
\end{align}
where $a$ and $f_i$ are assumed to belong to the functional class $\mathfrak{C}$. The linearized equations read 
\besubeqs
\begin{align}
 da-\qcommut{A^0}{a}&=u^i \star  F^0_i\,, &\delta a&=d\epsilon-\qcommut{A^0}{\epsilon}+ \lambda^i \star F^0_i\,,\\
 df_i-[A^0,f_i]_\star-[a,F^0_i]_\star&= u^j_i \star F^0_j\,, &\delta f_i&=\qcommut{\epsilon}{F^0_i}+ \lambda^j_i \star F^0_j\,,\\ \qcommut{F^0_i}{f_j}-(i\leftrightarrow j)-C_{ij}^k f_k&=  u^k_{ij}\star F^0_k\,.
\end{align} 
\esubeqs
We choose $\Pi a=a$, $\Pi f_i=f_i$ as a legitimate gauge condition. Just like in $X,P$-space the residual gauge symmetry $\delta f_i=\qcommut{\epsilon}{F^0_i}$  can be employed to gauge away  $f_0,f_-$ in $Y,P$-space. The only subtlety is that now both $f_i$ and $\epsilon$ are traceless and to see this one needs extra technique (see \cite{Barnich:2006pc} for details). 

With $f_0,f_-$ set to zero the equations for $a,f_+$ imply
\besubeqs
\label{intermed}
\begin{align}
%
%\begin{aligned}
da&=\Pi\big(\qcommut{A^0}{a}\big)\,, \label{inter1}\\
df_+&=\Pi\big(\qcommut{A^0}{f_+}\big)+(P\cdot \d_Y) a \,,\label{inter2}\\
(Y+V)\cdot \dl{P} a&=\big(P\cdot\dl{P}-(Y+V)\cdot \dl{Y}\big)a=0\,,\label{inter3}\\
(Y+V)\cdot \dl{P} f_+&=\big(P\cdot\dl{P}-(Y+V)\cdot \dl{Y}-2\big)f_+=0\label{inter4}\,.
%\end{aligned}
\end{align}
\esubeqs
Note that for $A^0$ a flat $\mathfrak{so}(d,2)$-connection, the $\Pi$ projector is not needed and equations \eqref{intermed} are known~\cite{Barnich:2006pc} to describe Fronsdal fields on AdS space.

The equations~\eqref{intermed} can be reduced even further down to the unfolded form using a suitable generalization of the reduction put forward in~\cite{Barnich:2006pc}. Indeed, the procedure is purely algebraic and allows to eliminate all the components in $f_+$ which are in the image of the operator $P\cdot\dl{Y}$ and all the components in $a$ which are not in the kernel of $P\cdot\dl{Y}$. The remaining fields $\bar{f}$ and $\bar a$ satisfy  \eqref{inter3}, \eqref{inter4} and are such that $(P\cdot \d_Y) \bar a=0$ and $\bar{f}$ parametrizes the quotient $f_+\sim f_++(P\cdot\d_Y) \epsilon$. The reduced equations have the following structure
\begin{align}\label{almostunfld}
\begin{aligned}
d\bar a&=\Pi\big(\,\qcommut{A^0}{\bar a}\big)+\mu(A^0,A^0,\bar{f})\,,\\
d\bar{f}&=\Pi\big(\,\qcommut{A^0}{\bar{f}}\big)\,,
\end{aligned}
\end{align}
where $\mu(A^0,A^0,\bar{f})$ is a trilinear form.

The structure of the equations~\eqref{intermed} and \eqref{almostunfld} becomes more clear if one reformulates them in terms of certain modules of the on-shell HS algebra. To this end let us consider the following two modules: 
\besubeqs
\begin{align}
    M^0&=\{f \in \mathfrak{C}\,:\, \qcommut{F^0_-}{f}=\qcommut{F^0_0}{f}=0\}\,,\\
    M^1&=\{f \in \mathfrak{C}\, :\, \qcommut{F^0_-}{f}=\qcommut{F^0_0}{f}-2f=0\}\,.
\end{align}
\esubeqs
Note that these are precisely the spaces \eqref{inter3} and \eqref{inter4} where $a$ and $f_+$ belong to. It is easy to see that for any $f\in M^{0,1}$ and $B$ from the on-shell HS algebra \eqref{onshellHSalg}, $B\star f$ and
$f\star B$ belong to $f\in M^{0,1}$ so that $f\in M^{0,1}$ are bimodules over the HS algebra seen as an associative algebra.

Furthermore, it is clear that the operator $\sdd  =\qcommut{\cdot}{F^0_+}=P\cdot\dl{Y}$ defines a map $M^0\to M^1$ and moreover $\sdd (B \star f)=B \star (\sdd f)$  for any $B$ in the HS algebra.  It follows, that both $m^0 \equiv \ker \sdd \subset M^0$ and $m^1\equiv \mathrm{coker}\, \sdd \subset M^1$ (i.e. the quotient of $M^1$ modulo $\im \sdd$) are also modules of the HS algebra as an associative algebra.

In what follows we consider $M^{0,1}$ and $m^{0,1}$ as modules of
the HS algebra seen as a Lie algebra, with the action being:
\begin{equation}
    \rho(B)f=\qcommut{B}{f}
\end{equation}
In particular, $M^{0,1}$ and $m^{0,1}$ are modules over $\mathfrak{so}(d,2)$ which is a Lie subalgebra of the HS algebra. It is now clear 
that equations~\eqref{intermed} and \eqref{almostunfld}
are, up to an extra term, nothing but covariant-constancy conditions with respect to the HS algebra connection $A_0$. 

Strictly speaking the above construction applies to the off-shell version of the algebra and modules. The on-shell version is obtained by requiring all the elements to be totally traceless and applying the projector $\Pi$ when necessary. The modules $m^{0,1}$ of the on-shell HS algebra are known in the literature as, respectively, the adjoint and the twisted-adjoint modules \cite{Vasiliev:1988sa} (also as, respectively, gauge and Weyl modules). The above realization of the modules originates from~\cite{Barnich:2006pc} (see also~\cite{Chekmenev:2015kzf}), where they were considered as $\mathfrak{so}(d,2)$-modules only.

System \eqref{almostunfld} describes propagation of free HS fields, encoded  in $\bar a$ and $\bar{f}$, over a background described by the flat connection $A^0$ taking values in the on-shell HS algebra. The most nontrivial part of the system comes from the term $\mu(A^0,A^0,\bar{f})$ which is trilinear in the fields and cannot be reduced to a product in the HS algebra. In the next section, we explain that this is a correct structure which is completely fixed by the HS algebra. Therefore, the parent form of the HS extension of the FG construction solves an important problem of how to make HS fields propagate on backgrounds that differ from pure AdS space. The derivation above was quite abstract and we do not aim at deriving the explicit form of $\mu(A^0,A^0,\bar{f})$ (its free approximation when $A^0$ is an $\mathfrak{so}(d,2)$ connection was discussed many times, see e.g. \cite{Vasiliev:2003ev,Bekaert:2005vh,Barnich:2006pc}). This is due to the fact that the explicit form of $\mu(A^0,A^0,\bar{f})$ can be changed by field redefinitions and it is difficult to compare with vertices in the usual weak-field expansion. 

Let us discuss if the on-shell conditions entering the full system \eqref{full-parent} considered over the flat vacuum $A^0,F^0_i$
can be extended beyond the linear order. The general feature of the formulations that are based on jet bundles is that the field space contains infinitely many auxiliary fields that, as a consequence of equations of motion, encode derivatives of the fields of arbitrarily high order. The advantage is that interaction terms can then be written in an algebraic form. However, this does not come for free and non-linear expressions can easily contain infinitely many derivatives and make locality properties obscure. This problem becomes visible when nonlinearities are at least bilinear in fluctuations $\bar{f}$, i.e. are of order $\mathcal{O}(\bar{f}^2)$. This is due to the fact that the HS algebra has a well-defined grading that is mapped to polynomial degree in $Y+V$, $P$. Therefore, expressions of order $\mathcal{O}(\bar{a}\bar{f})$ or $\mathcal{O}(\bar{a}\bar{a}\bar{f})$ are always local once we fix the spins (i.e. homogeneity in $P$) in $\bar{a}$ and $\bar{f}$. However, expressions of order $\mathcal{O}(\bar{f}^2)$ can have unbounded number of derivatives.\footnote{Indeed, $\bar{a}$ contains a finite number of derivatives per spin and $\bar{f}$ contains derivatives of unbounded order. Looking at the possible contribution of the interaction vertices to a given equation we see that infinite sums over derivatives require at least two $\bar{f}$'s on the right-hand-side.} Such non-localities arise at the quartic order in weak-field expansion \cite{Bekaert:2015tva,Sleight:2017pcz}, i.e. $\mathcal{O}(\bar{a}\bar{f}^3)$ $\mathcal{O}(\bar{a}\bar{a}\bar{f}^2)$ in the parent formulation~\eqref{full-parent}. In the framework offered by the nonlinear system \eqref{full-parent} understood perturbatively over the flat vacuum $A^0,F^0_i$  the locality problem manifest itself in that the functional class $\mathfrak{C}$ is not closed under star-multiplication. Although this does not affect the linearized system~\eqref{intermed}, at higher orders either the functional class or even the system itself has to be amended.

%%%%%%%%%%%%%%%%%%%%%%%%%%%%%%%%%%%%%%%%%%%%%%%%%%%%%%%%%%%%%%%%
\subsection{Relation to Unfolded Equations}
%%%%%%%%%%%%%%%%%%%%%%%%%%%%%%%%%%%%%%%%%%%%%%%%%%%%%%%%%%%%%%%%
The system \eqref{almostunfld} can also be understood as a specific free differential algebra \cite{Sullivan77,D'Auria:1982pm, Nieuwenhuizen:1982zf}, unfolded equations \cite{Vasiliev:1988sa} or AKSZ sigma model \cite{Alexandrov:1995kv} associated to a certain target-space $Q$-manifold (see also~\cite{Barnich:2005ru}). 
The underlying $Q$-manifold can be directly related to the deformation procedure that is relevant for higher-spin theories~\cite{Vasiliev:1988sa}. Our aim is to show how  \eqref{almostunfld} arises. The field content, or the coordinates of the $Q$-manifold, consists of a connection of the on-shell HS algebra $\omega$ and a zero-form $C$ that takes values in the on-shell HS algebra as well. The deformation procedure starts with the flatness condition for $\omega$ and a covariant constancy condition for $C$:
\besubeqs\label{unfoldedstart}
\begin{align}
&d\omega=\omega\ast\omega\,,\\
&dC=\omega\ast C-C\ast \pi(\omega)\,,
\end{align}
\esubeqs
where the automorphism $\pi$ is induced by an automorphism of $\mathfrak{so}(d,2)$ that flips the sign of the transvection generators and leaves Lorentz generators intact. Note that in contrast to the previous section $\omega$ and $C$ take values in the on-shell HS algebra from the very beginning and the associative product is denoted $\ast$. As we discussed in the previous section the twisted-adjoint covariant constancy equation for $C$ can be systematically derived from the parent system~\eqref{intermed}, which in turn arises from the HS extended FG theory upon factorization.

System \eqref{unfoldedstart} involves a HS-flat background $\omega$ and a linear fluctuation $C$. When %linearized
considered over AdS background, given by $\omega$ belonging to the $\mathfrak{so}(d,2)$ subalgebra, the equation for $C$  describes a scalar field and $s=1,2,3,...$ massless fields in terms of their gauge-invariant field strengths contained in $C$: Faraday tensor, Weyl tensor and higher-spin generalizations thereof \cite{Vasiliev:1988sa}.

The deformation procedure has $C$ as an expansion parameter. Therefore, the first order deformation  of \eqref{unfoldedstart} involves a vertex that violates the flatness of $\omega$: 
\besubeqs
\label{linearC}
\begin{align}
&d\omega=\omega\ast\omega+\mathcal{V}(\omega,\omega,C)+\mathcal{O}(C^2)\,,\\
&dC=\omega\ast C-C\ast \pi(\omega)+\mathcal{O}(C^2)\,.
\end{align}
\esubeqs
The consistency conditions on $\mathcal{V}$ follow from $d\,d\equiv0$. The solution for $\mathcal{V}$ can be cast into the form \cite{Sharapov:2017yde}\footnote{A crucial assumption needed to reduce a complicated problem of Chevalley-Eilenberg cohomology of higher-spin algebra to a much simpler problem of Hochschild cohomology is to assume that higher-spin theory should allow for Yang-Mills gaugings $u(M)$ for any $M$. This assumption is justified by AdS/CFT correspondence where higher-spin theories are dual to free CFT's and it is always possible to take a number of free fields as to have $u(M)$ (times higher-spin algebra) as a global symmetry on the CFT side and hence HS algebra tensored with matrices is a gauge symmetry on the AdS side. Having this matrix factor allows one to reduce 
the problem \cite{Sharapov:2017yde} to the Hochschild cohomology.}  
\begin{align}\label{firstorder}
\mathcal{V}(\omega,\omega,C)=\Phi(\omega,\omega)\ast \pi(C)\,,
\end{align}
where $\Phi(\bullet,\bullet)$ is a Hochschild two-cocycle of the higher-spin algebra in the twisted-adjoint representation:
\begin{align}\label{sptwoHochschildA}
    a\ast \Phi(b,c)+\Phi(a\ast b,c)-\Phi(a,b\ast c) +\Phi(a,b)\ast \pi(c)=0\,.
\end{align}
The deformation does not stop at $\mathcal{O}(C^2)$ order and higher orders are needed. For a large class of algebras it can be shown \cite{Sharapov:2017yde,Sharapov2017} that there are no obstructions at higher orders. Therefore, any consistent system that has \eqref{linearC} as the order-$\mathcal{O}(C)$ approximation can be completed to a solution to the full non-linear deformation problem.  The conclusion is that the full nonlinear system is determined by HS algebra, its twisted-adjoint module and the vertex $\mathcal{V}(\omega,\omega,C)$.

The system that describes fluctuations of HS fields over any HS-flat background  is obtained from \eqref{linearC} by taking any flat connection $A^0$ and expanding the system $\omega\rightarrow A^0+\omega$ to the first order in $C$ and $\omega$: 
\besubeqs\label{linfluct}
\begin{align}
dA^0&=A^0\ast A^0\,,\\
d\omega&=A^0 \ast \omega +\omega\ast  A^0+\mathcal{V}(A^0,A^0,C)\,,\\
dC&=A^0 \ast C-C\ast \pi( A^0)\,.
\end{align}
\esubeqs
These equations are fully consistent, gauge invariant and do not require any higher order corrections. They can be identified with the equations \eqref{Flaths}, \eqref{almostunfld} obtained as a linearization of the HS-extension of the FG-construction \eqref{full-parent}.

To sum up, we observe that all the structures governing HS theories within the unfolded approach are already present in the linearization over a HS-flat background. In its turn this system and the respective structures can be extracted from the nonlinear system \eqref{full-parent}, which in turn can be related to the HS-extension of the FG construction. Equations \eqref{Flaths}, \eqref{almostunfld} from the previous section are exactly of this form. Fluctuations over a HS-flat background seems to be the farthest one can get without facing the locality problem in this approach.

%%%%%%%%%%%%%%%%%%%%%%%%%%%%%%%%%%%%%%%%%%%%%%%%%%%%%%%%%%%%%%%%%
\section{Conclusions and Discussion}
\label{sec:discussion}
%%%%%%%%%%%%%%%%%%%%%%%%%%%%%%%%%%%%%%%%%%%%%%%%%%%%%%%%%%%%%%%%%

In this work we have shown that the $\mathfrak{sp}(2)$-constraints on the ambient phase space are equivalent to the off-shell Fefferman-Graham theory, if the three constraints are of degree $0,1,2$ in the momentum. The HS extension then follows by letting the three constraints to have arbitrary powers of momenta. The HS extended  $\mathfrak{sp}(2)$-system has already been studied in the past, both in the context of bulk HS theories~\cite{Grigoriev:2012xg} (see also~\cite{Bars:2001um,Grigoriev:2006tt}) and conformal HS fields on the boundary~\cite{Bonezzi:2010jr,Bekaert:2013zya,Alkalaev:2014nsa}. 

It is more convenient to analyze the equations after reformulating the system in parent form, i.e. by moving the ambient space-time to the fiber and introducing an extra gauge field associated with ambient diffeomorphisms. In so doing the original space time coordinates $X^A$ are promoted to the components of the compensator field $V^A$. In particular, the bulk theory can now be formulated in terms of intrinsic geometry of AdS space by pulling back the ambient parent system to $d+1$ dimensions and setting $V^2=-1$.

Although it was known that the linearized parent formulation of $\mathfrak{sp}(2)$-system can be put on-shell by supplementing the system with extra conditions, thereby giving rise to the infinite multiplet of on-shell Fronsdal fields, it was not clear how to implement this beyond the linearized approximation over the AdS vacuum. In this work, we propose a procedure that allows to go one step further such approximation.
More precisely, we consider a factorized version of the parent $\mathfrak{sp}(2)$-system. Although with the simplest functional class the system is empty when considered over AdS background, we have succeeded to find another functional class (in auxiliary space) such that the modified system admits HS-flat vacuum solution and the respective linearized system is nonempty and properly describes propagation of HS fields over a HS-flat background. Even though the extension to higher orders still remains an open problem, this linearized system already contains all the structures determining the Vasiliev equations.

It has been known for many years that free massless HS fields cannot be put on nontrivial gravitational backgrounds as it leads to breakdown of gauge invariance. However, HS gauge  fields \textit{can} propagate on nontrivial backgrounds that have other higher-spin gauge fields turned on --- backgrounds described by a flat connection of a HS algebra.\footnote{In three dimensions HS fields do not have propagating degrees of freedom but HS-flat backgrounds were found to contain many interesting solutions, e.g. black holes and conical defects.} The resulting equations have a clear algebraic meaning of flatness condition deformed by a Hochschild two-cocycle of the relevant HS algebra. This probes the structure of interactions, even though it is hard to directly make a link to the vertices in the weak field expansion over anti-de Sitter space. 

It can be argued~\cite{Sharapov:2017yde} that, at least within the formal deformation procedure, a HS algebra and its Hochschild cocycle determine the full non-linear completion so that the knowledge of free fluctuations over sufficiently general backgrounds is enough to recover the full structure of interactions. We showed that all this information is already present in the factorized version of the parent $\mathfrak{sp}(2)$-system.
The problem of putting the system on-shell at higher orders is clearly related to the subtlety of higher-spin interactions in anti-de Sitter space: the interactions are known to be non-local starting from the quartic order \cite{Bekaert:2015tva,Sleight:2017pcz,Ponomarev:2017qab} and the precise characterization of the appropriate functional class is not yet known. 

The factorized version of the parent $\mathfrak{sp}(2)$-system employed in the paper also has a natural interpretation in the context of CHS theories. Indeed, the same system considered on the boundary, 
by setting $V^2=0$,
describes nonlinear CHS 
fields at the off-shell level. More specifically, it is equivalent (at least perturbatively) to the system from~\cite{Segal:2002gd} underlying the nonlinear theory of CHS fields.

It is not surprising that that the system describing off-shell conformal HS fields on the boundary has something to do with
on-shell HS fields in the bulk. Indeed, the former are the boundary  values of the later while in the ambient space formulation bulk fields and their boundary values are typically described by one and the same ambient system considered, respectively, around the hyperboloid and around the hypercone (see e.g. ~\cite{Bekaert:2012ux,Bekaert:2013zya} for a parent formulation).  Therefore, the parent $\mathfrak{sp}(2)$-system provides a direct link between the symmetries of the effective action of the scalar field on CHS background and HS gravity in the bulk.

From this perspective, the approach advocated in this paper can be viewed as a purely classical version of the holographic reconstruction (see e.g.  \cite{deHaro:2000vlm,Bekaert:2015tva,Sleight:2017pcz} for somewhat related  approaches). Indeed, the candidate system describing nonlinear on-shell HS fields in the bulk is obtained by pulling to the bulk the boundary system describing off-shell conformal fields. It is a remarkable feature of the ambient space formulation that it does not only allows one to go from bulk to boundary but also to reverse the procedure in order to reconstruct bulk theory from the boundary values. Strictly speaking, to make it work one also employs the parent formulation as to fine-tune the system by picking specific functional classes in the auxiliary space.

Possible generalizations and extensions of the above construction involve a number of cases: theories of partially-massless higher-spin fields \cite{Bekaert:2013zya,Alkalaev:2014nsa, Brust:2016zns}, which result from the same $\mathfrak{sp}(2)$-system but making use of GJMS operators \cite{GJMS} for the higher-order singleton and a slightly different factorization procedure \cite{Alkalaev:2014nsa,Joung:2015jza}. It would be also interesting to consider other algebras of constraints. For example, $\mathfrak{osp}(1|2)$-constraints should describe the yet-unknown Type-B theory that is dual to free massless fermions and Gross-Neveu model. This problem was recently discussed in \cite{Bonezzi:2014nua,Bonezzi:2017hxd}. 

Finally, let us note that there exists an alternative approach to described (HS) fields in $AdS_{d+1}$ in terms of $\mathfrak{sp}(2)$-system in $(d+3)$-dimensional ambient space. In so doing one needs an additional ingredient, the scale tractor, that breaks the conformal invariance of the tractor formulation \cite{BEG,Gover:2011rz}.

%%%%%%%%%%%%%%%%%%%%%%%%%%%%%%%%%%%%%%%%%%%%%%%%%%%%%%%%%%%%%%%%%
\section*{Acknowledgments}
\label{sec:Aknowledgements}
%%%%%%%%%%%%%%%%%%%%%%%%%%%%%%%%%%%%%%%%%%%%%%%%%%%%%%%%%%%%%%%%%

We thank the organizers of the very stimulating workshop on ``Higher-spin gauge theories'' (26-28 of April, 2017, Mons, Belgium) for providing us the opportunity to present this work in a warm atmosphere. This research was supported by the Russian Science Foundation grant 14-42-00047 in association with the Lebedev Physical Institute.

%%%%%%%%%%%%%%%%%%%%%%%%%%%%%%%%%%%%%%%%%%%%%%%%%%%%%%%%%%%%%%%%
\begin{appendix}
\renewcommand{\thesection}{\Alph{section}}
\renewcommand{\theequation}{\Alph{section}.\arabic{equation}}
\setcounter{equation}{0}\setcounter{section}{0}
%%%%%%%%%%%%%%%%%%%%%%%%%%%%%%%%%%%%%%%%%%%%%%%%%%%%%%%%%%%%%%%%

%%%%%%%%%%%%%%%%%%%%%%%%%%%%%%%%%%%%%%%%%%%%%%%%%%%%%%%%%%%%%%%%
\section{FG Ambient Construction as $\mathfrak{sp}(2)$ Algebra of Constraints}
\label{app:xav}
%%%%%%%%%%%%%%%%%%%%%%%%%%%%%%%%%%%%%%%%%%%%%%%%%%%%%%%%%%%%%%%%

%\newpage

\subsection{Klein Flat Ambient Model}\label{Klein}

In modern conformal field theory and its holographic dual interpretation, a celebrated technique for performing computations is the ambient formalism. Its principle goes back to Dirac's cone reformulation of conformal fields and their wave equations \cite{Dirac:1936fq} corresponding to the case of flat Lorentzian conformal geometry. 

In turn, Dirac's approach goes back to Klein's model of flat Euclidean $d$-dimensional conformal geometry, that is the round sphere ${\mathbb S}^d$ with conformal isometry algebra $\mathfrak{so}(d+1,1)$. The main idea of the ambient approach is to make conformal symmetry manifest via an embedding of this geometry inside an ``ambient'' ($d+2$)-dimensional Minkowski space ${\mathbb R}^{d+1,1}$. The ($d+1$)-dimensional upper null cone $N\subset {\mathbb R}^{d+1,1}$, generated by light-like rays through the origin, plays a crucial role.
The conformal sphere ${\mathbb S}^d$ is realized as the projective future light-cone ${\mathbb P}N:=N/{\mathbb R}^+$ (with ${\mathbb R}^+$ the multiplicative group of positive real numbers) inside the (d+1)-dimensional projective ambient space ${\mathbb P}{\mathbb R}^{d+1,1}:={\mathbb R}^{d+1,1}/{\mathbb R}^+$. The interior of the projective future light-cone ${\mathbb P}N\subset{\mathbb P}{\mathbb R}^{d+1,1}$ is a (d+1)-dimensional hyperbolic ball ${\mathbb H}_{d+1}\cong{\mathbb B}^{d+1}$, of which the d-dimensional sphere ${\mathbb S}^d$ is the conformal boundary.

\subsection{Fefferman-Graham Ambient Construction}\label{FGapp}

In 1985, Fefferman and Graham generalized the ambient construction of Klein to \textit{curved} conformal geometry (of any signature) \cite{FG}.

\vspace{1mm}
\noindent\textbf{Conformal space:} The basic data of conformal geometry is a manifold $M$ and a conformal metric $[g_{ab}]$ on $M$, i.e. an equivalence class of Riemannian metrics for the equivalence relation 
\begin{equation}
\tilde{g}_{ab}=\Omega^2 g_{ab}\sim g_{ab}
\end{equation}
with $\Omega$ a nowhere vanishing function on $M$ (which we will assume strictly positive $\Omega>0$).

\noindent\textit{Example:} In the case of flat Euclidean conformal geometry, $M={\mathbb S}^d$ and $g_{ab}$ is the standard metric of the unit sphere (which is conformally flat since the flat metric $\delta_{ab}$ belongs to the same equivalence class $[g_{ab}]$).

\vspace{1mm}

\noindent\textbf{Metric bundle:} The first step in the construction of Fefferman and Graham is the introduction of a principal ${\mathbb R}^+$-bundle $N$ over $M$, 
%with projection map $\pi:N\to M$,
 whose fiber at a point (of coordinates $x^a$ with $a=1,\cdots,d$) is the collection of values at this point of all representatives $\tilde{g}_{ab}(x)=\Omega^2(x) g_{ab}(x)$ inside the conformal class. The base manifold is recovered as the quotient $M=N/{\mathbb R}^+$.
One may take ($x^a,t$) as local coordinates on $N$ with $t:=\Omega(x)$. This bundle is called the \textit{metric bundle} because its sections are the representatives $\tilde{g}_{ab}$ of the given conformal class $[g_{ab}]$. The fundamental vector field of the principal ${\mathbb R}^+$-bundle $N$ will be denoted $v$ ($=t\partial_t$ in local coordinates). The conformal class $[g_{ab}]$ of metrics defines a degenerate metric on $N$ by pullback along the projection map $\pi:N\to M$\,,
\begin{equation}\label{degmetr}
ds_N^2\,=\,t^2\,g_{ab}(x)\,dx^adx^b\,.
\end{equation}
This metric is homogeneous of degree 2 under the action of ${\mathbb R}^+$ on $N$.
However, this metric is degenerate, it annihilates for instance the fundamental vector field $v$.

\noindent\textit{Example:} In the case of flat Euclidean conformal geometry the metric bundle $N$, as it was introduced, has the topology of a cylinder ${\mathbb S}^d\times{\mathbb R}^+$, which can be interpreted as an upper cone (but it can be extended to a complete cone if we extend the range of values of $t$ to all real numbers). Together with its degenerate metric, it is indeed a null cone, as mentioned in the section \bref{Klein}.

\vspace{1mm}

\noindent\textbf{Ambient space:} The second step in the construction of Fefferman and Graham is to embed the ($d+1$)-dimensional manifold $N$ inside a ``slightly thicker'' ($d+2$)-dimensional \textit{ambient space}, e.g. $\tilde{N}=N\times I$, where $I\subset \mathbb R$ is an open interval around zero. 
The natural extension to $\tilde N$ of the fundamental vector field $v$ on $N$ is a vector field $V$ on $\tilde N$ which one might call the \textit{homothety vector field}.
A \textit{defining function} $\rho$ is a function on $\tilde{N}$ with homogeneity degree zero under the homothety vector field $V$, and such that $\rho=0$ but $d\rho\neq 0$ on $N$.
The ambient space $\tilde N$ is endowed with local coordinates $Y^M$ = ($x^a,t,\rho$) where $\rho$ is a defining function. An \textit{ambient metric} $G_{MN}$ is a metric on ambient space $\tilde N$ such that:
\begin{itemize}
	\item its signature has one more timelike and one more spacelike direction with respect to $g_{ab}$,
  \item it is homogeneous of degree two with respect to the homotheties: ${\cal L}_V G_{MN}=2\,G_{MN}$,
	\item it is an extension to $\tilde N$ of the degenerate metric \eqref{degmetr} on $N$,
	\item the one-form $V_M=G_{MN}V^N$
	 is closed.
\end{itemize}
There exists a local choice of coordinates such that the ambient metric reads \cite{Fefferman:2007rka}
\begin{equation}\label{ambmetr}
ds_{\tilde{N}}^2\,=\,t^2\,g_{ab}(x)\,dx^adx^b\,+\,2\,\rho\,dt^2\,+\,2\,t\,dt\,d\rho
\end{equation}
and the vector field $\partial/\partial\rho$ is geodesic. The square of the homothety vector field is proportional to the defining function $\rho$ since, from \eqref{ambmetr}, one has $G_{MN}V^MV^N=t^2\rho$\,. Taking \eqref{ambmetr} into account gives 
$V_M dY^M=2\rho\, t\,dt+t^2\,d\rho=d(t^2\rho)$ in  local coordinates. 

\noindent Up to the well-known subtleties in the holographic reconstruction related to conformal anomalies, the ambient metric is essentially uniquely specified (in an infinitesimally thick neighborhood around $N$ or, more precisely, as a formal power series in the variable $\rho$) if one further requires $G_{MN}$ to be Ricci flat: $R_{MN}=0$. 

\noindent\textit{Example:} In the case of flat Euclidean conformal geometry, the ambient space is the Minkowski spacetime $\tilde{N}={\mathbb R}^{d+1,1}$ with ambient metric $\eta_{MN}$ = diag(-1,+1,...,+1) in the Cartesian coordinates $X^M$ with $M=0,1,\cdots,d+1$. The homothety vector field reads $V=X^M\partial_M$ and the metric bundle $N$ is embedded as the light-cone through the origin $V^2=0\Leftrightarrow\eta_{MN}X^MX^N=0$. The relation between the Cartesian coordinates $X^M$ and the FG coordinates $Y^M$ = ($x^a,t,\rho$) is as follows: $X^a=t\,x^a$, $X^0-X^{d+1}=t$ and $X^0+X^{d+1}=t\,(\,\delta_{ab}x^ax^b-2\rho\,)$\,. In such case, $ds_{\tilde{N}}^2=\eta_{MN}dX^MdX^N$ reproduces \eqref{ambmetr} with $g_{ab}=\delta_{ab}\,$.

\vspace{1mm}

\noindent\textbf{Bulk:} The third step in the construction of Fefferman and Graham provided its holographic interpretation, cherished by theoretical physicists.
The \textit{bulk space} is the (d+1)-dimensional manifold $\tilde{M}:=(\tilde{N}-N)/{\mathbb R}^+$. One may take ($x^m,\rho$) as local coordinates on the bulk space $\tilde M$ which can be realized as the level hypersurface $t^2\rho$ = constant.
It is endowed with a metric $g_{\mu\nu}$ which has $[g_{ab}]$ as conformal class at conformal infinity (i.e. at $\rho=0$).
If the ambient metric $G_{MN}$ is Ricci flat ($R_{MN}=0$), then the bulk metric is Einstein ($R_{\mu\nu}\,=\,\frac1{d+1}\,R\,g_{\mu\nu}$).

\noindent\textit{Example:} In the case of flat Euclidean conformal geometry, to be more precise the ($d+2$)-dimensional ambient space is the interior future light-cone $\tilde{N}$ inside the Minkowski spacetime ${\mathbb R}^{d+1,1}$. Then the bulk space is the ($d+1$)-dimensional hyperbolic space ${\mathbb H}_{d+1}$ realized as one sheet of a two-sheeted hyperboloid $\eta_{MN}X^MX^N=-1$ of time-like vectors of constant square in ambient space. This corresponds to the level hypersurface $t^2\rho=-\tfrac12$ in the FG coordinates. Up to the change of coordinate $z=1/\,t$\,, this leads to the standard form $ds_{\tilde{M}}^2\,=\,z^{-2}\,(\,\delta_{ab}\,dx^adx^b\,+\,dz^2\,)$ of the hyperbolic space metric in Poincar\'e coordinates.

\subsection{Properties of the Ambient Metric and of the Homothety Vector Field}

The two main ingredients of Fefferman-Graham construction are the ambient metric $G_{AB}$ and the homothety vector field $V^A$. They are closely related to each other due to the two following properties:
\begin{itemize}
	\item[(I)] The ambient metric is of homogeneity degree two with respect to the homothety vector field: 
\begin{equation}\label{hometr}
{\cal L}_V G_{AB}\,=\,2\,G_{AB}\,.
\end{equation}
	\item[(II)] The homothety one-form is closed:
\begin{equation}\label{closed}
\partial_{[A}V_{B]}\,=\,0\,.
\end{equation}
\end{itemize}
In particular, the property (II) implies that, locally, $V_A=\partial_A f$. In particular, the homothety vector field $V^A$ is hypersurface orthogonal to the level surfaces $f$ = constant.

\vspace{3mm}

\noindent The properties (I) and (II) are equivalent to the following useful property:
\begin{itemize}
	\item[(III)] The ambient metric is equal to the covariant derivative of the homothety one-form: 
\begin{equation}\label{ambmetrhomvect}
G_{AB}\,=\,\nabla_AV_B\,=\,\nabla_BV_A\,.
\end{equation}
Equivalently,
\begin{equation}
\nabla_AV^B\,=\,\delta_A^B\,.
\end{equation}
\end{itemize}
\textit{Proof:} While the relation \eqref{hometr} is equivalent to
\begin{equation}\label{hometr'}
\nabla_AV_B\,+\,\nabla_BV_A\,=\,2\,G_{AB}\,,
\end{equation}
the relation \eqref{closed} is equivalent to $\nabla_{[A}V_{B]}=0$, i.e.
\begin{equation}\label{closed'}
\nabla_{A}V_{B}\,-\,\nabla_{B}V_{A}\,=\,0\,.
\end{equation}
Summing the equations \eqref{hometr'} and \eqref{closed'} gives \eqref{ambmetrhomvect}. \qed

\vspace{3mm}

\noindent In turn, the property (III) implies the following corollary:
\begin{itemize}
	\item[(IV)] The homothety one-form is equal to half the gradient of the homothety vector field squared:
\begin{equation}\label{gradient}
V_A\,=\,\partial_A\Big(\frac{V^2}2\Big)\,.
\end{equation}
\textit{Proof:} Contracting $G_{AB}=\nabla_AV_B$ with $V^B$ gives the relation $V_A=V^B\nabla_AV_B$ which implies \eqref{gradient}. \qed
\end{itemize}
\noindent In particular, the homothety vector field is hypersurface orthogonal to the level surfaces $V^2$ = constant.

\subsection{Hypersurface Orthogonality and Homogeneity as $\mathfrak{sp}(2)$ Algebra}\label{hyporth}

Consider the cotangent bundle $T^*\tilde{N}$ of the ambient space. Local coordinates on $T^*\tilde{N}$ read ($X^M,P_N$) and the canonical Poisson bracket on the algebra of functions on $T^*\tilde{N}$ is such that $\{X^M,P_N\}=\delta^M_N$. 
In the main text, we need the relation between the Fefferman-Graham ambient construction and the $\mathfrak{sp}(2)$ algebra of constraints.

Consider as data, three Hamiltonian constraints which are respectively quadratic, linear, independent, of the momenta:
\begin{eqnarray}
F_+&=&\frac12 P_M P_N G^{MN}(X)\label{cstr1}\\
F_0&=&P_M V^M(X)\label{cstr2}\\
F_-&=&F(X)\label{cstr3}
\end{eqnarray}
The first Hamiltonian constraint is quadratic in the momenta, thus its coefficients define a covariant symmetric tensor $G^{MN}(X)$, which can be interpreted as an inverse metric if it is nondegenerate. This fact %(degree two in the momenta and invertibility)
will be called the \textit{nondegeneracy condition}. The second Hamiltonian constraint is linear in the momenta, thence its coefficients define a vector field $V^M(X)$. The third Hamiltonian constraint is a scalar field $F(X)$.

\noindent\textbf{Proposition:}
Under the assumption of nondegeneracy, the property that the three functions 
\eqref{cstr1}-\eqref{cstr2}
on the ambient phase space form the $\mathfrak{sp}(2)$ algebra 
\begin{equation}\label{sp2constrs}
\{F_+,F_-\}=F_0\,,\quad \{F_0,F_+\}=+\,2\,F_+\,,\quad \{F_0,F_-\}=\,-\,2\,F_-\,,
\end{equation}
under the Poisson bracket is equivalent to the fact that 
\begin{itemize}
    \item the scalar field is equal to $F(X)=-\frac12\,V^M(X)\,G_{MN}(X)\,V^N(X)$,
    \item the symmetric tensor field $G_{MN}(X)$ and the vector field $V^M(X)$ obey the properties (I)-(IV).
\end{itemize}
 
\noindent\textit{Proof:} 
Since $h=P\cdot V(X)$ is linear in the momenta, the adjoint action of the constraint $h$ via the Poisson bracket is related to the Lie derivative along the homothety vector field, $\{h,\cdot\}=-{\cal L}_V$\,. Therefore, the second relation in \eqref{sp2constr} is equivalent to the condition ${\cal L}_V G^{MN}\,=\,-\,2\,G^{MN}\,\Leftrightarrow\,{\cal L}_V G_{MN}\,=\,2\,G_{MN}$, i.e. it is equivalent to \eqref{hometr}. The first relation in \eqref{sp2constrs} gives the relation $\partial_M F\,=\,-\,V_M$, i.e. it implies the property (II) in the section \bref{offshellFG}. Together with property (I), it implies the properties (III)-(IV) of the section \bref{offshellFG}. In particular, the relation \eqref{gradient} implies
$\,\partial_M F=-\,\partial_M\big(\frac{V^2}2\big)$. This leads to $F(X)=C-\frac12\,V^M(X)G_{MN}(X)V^N(X)$, where $C\in\mathbb R$ an arbitrary constant which is enforced to vanish by the third relation in \eqref{sp2constrs}.
Then, the third relation in \eqref{sp2constrs} is equivalent to the homogeneity property ${\cal L}_V G_{MN}\,=\,2\,G_{MN}$. \qed

\section{Covariant Derivatives}
\label{covder}

Suppose we are given with the ambient metric $g_{MN}(X)$ and the compensator $V^M(X)$ such that $\nabla_MV_N=G_{MN}$, where $\nabla$
denotes the covariant derivative determined by the Levi-Civita connection.

Introduce the covariant derivative acting on functions in $X,P$ as follows:
$$
D_M f(X,P):=\d_M+\Gamma_{MN}^RP_R\dl{P_N}
$$
so that  e.g. 
$$
D_M (V^NP_N)=(\nabla_M V^N)P_N\,.
$$
It is easy to check that the Poisson bracket of functions in $X,P$ can be written in terms of the covariant derivative:
$$
\pb{F}{G}:=\ddl{F}{X^M}\ddl{G}{P_M}-\ddl{F}{P_M}\ddl{G}{X^M}=D_M F\ddl{G}{P_M}-\ddl{F}{P_M} D_M G\,.
$$
For instance, for the adjoint action of $\mathfrak{sp}(2)$ generators we have:
\begin{eqnarray}
\pb{V^MP_M}{f}&=&(\d_N V^M) P_M\left(\dl{P_N}f\right)-V^M\d_M f
\nonumber\\
&=&(\nabla_N V^M)P_M\left(\dl{P_N}f\right)-\Gamma_{NR}^MV^RP_M\left(\dl{P_N}f\right)-V^M\d_M f\nonumber\\
&=&
(\nabla_N V^M)P_M\left(\dl{P_N}f\right)-V^AD_Af\nonumber\\
&=&\big(P_M\dl{P_M}-V^AD_A\big)f\,.
\end{eqnarray}
Similarly for the other generators.

%%%%%%%%%%%%%%%%%%%%%%%%%%%%%%%%%%%%%%%%%%%%%%%%%%%%%%%%%%%%%%%%%%%%%%%%%%
%%%%%%%%%%%%%%%%%%%%%% APPENDIX C %%%%%%%%%%%%%%%%%%%%%%%%%%%%%%%%%%%%%%%%
\section{Off-shell vs On-shell, Boundary vs Bulk}
\label{sec:off/on}

The different physical interpretations of the FG ambient construction, mentioned in sections \bref{offshellFG} and \bref{onshellFG}, can be illustrated in the simple case of a scalar field, which is an inspiring toy model.\footnote{For more details on the ambient formulation of the scalar field and its holographic interpretation (in the case of the flat ambient space) see \textit{e.g.} \cite{Bekaert:2012vt,Bekaert:2013zya}.} This simple example is actually of interest on its own, since it corresponds to the first-quantization of the $\mathfrak{sp}(2)$ Hamiltonian constraints. To be more precise, we consider an ambient scalar field in a background of off-shell FG theory.

\paragraph{Off-shell boundary scalar field.}

The null cone $V^2(X)=0$ quotiented by the integral lines of the vector field $V=V^A\partial_A$ is a $d$-dimensional conformal space.
A scalar primary conformal field of conformal weight $\Delta$ on this $d$-dimensional conformal space can be lifted uniquely to a scalar field on the null cone $V^2=0$ with homogeneity degree $-\Delta$\,. However, the latter does not determine a unique scalar field $\Phi(X)$ in the vicinity of the null cone, but only up to the following equivalence relation $\Phi\sim \Phi\, + \,V^2 \,\lambda$\,.
This can be summarized by saying that an off-shell conformal scalar field in $d$ dimensions is equivalently described by an ambient scalar field obeying to the following set of one equation and one equivalence relation:
\begin{equation}\label{C1}
    (V^A \d_A+\Delta)\Phi(X)=0\,, \qquad  \Phi(X)\sim \Phi(X) + V^2(X) \,\lambda(X)\,.
\end{equation}
{The consistency of the above equations amounts to the fact that the operators $V^A\d_A$ and $V^2$ form a Lie algebra (the lower-triangular subalgebra of $\mathfrak{sp}(2)$)}.  

\noindent
\textit{Spin-two analogue:} Off-shell conformal gravity determines the value of the ambient metric on the null cone. The ambient metric in off-shell FG theory has homogeneity degree two and its extension beyond the null cone is completely undetermined.

\paragraph{On-shell bulk scalar field.}

An on-shell scalar field on the $d+1$ dimensional level manifold $V^2(X)=-1$, i.e. a scalar field obeying to  Klein-Gordon equation in the bulk, can be lifted uniquely to a scalar field $\Phi(X)$ on the region $V^2<0$ of the ambient space. The lift is unique if the scalar field is of fixed homogeneity degree, say $-\Delta\in\mathbb R$, along the homothety vector field $V$, i.e. ${\cal L}_V\Phi=-\Delta\,\Phi\,$. More precisely, the ambient scalar field obeys the following two equations:
\begin{equation}\label{C2}
 (V^A \d_A+\Delta)\Phi(X)=0\,, \qquad  \nabla^2 \Phi(X)=0\,,
%    \d_X\cdot\d_X \Phi& =0 &&  (X\cdot \d_X+\Delta)\Phi=0
\end{equation}
where $\nabla^2=G^{AB}\nabla_A\nabla_B$ is the Laplacian for the ambient metric. The consistency of the two conditions in \eqref{C2} can be checked by using $\nabla_A V ^B=\delta_A^B$ and its consequence $R_{AB|CD}V^D=0\,$. Moreover, these two operators form  the upper-triangular subalgebra of $\mathfrak{sp}(2)$. 

When the scalar field $\Phi(X)$ admits a regular extension to the whole region $V^2\leqslant 0$ of the ambient space, then its restriction to the null cone $V^2=0$ corresponds to the asymptotic boundary data of the on-shell bulk scalar field, with scaling behavior prescribed by $\Delta$ (see e.g. \cite{Bekaert:2012vt} for more details). To be more precise, two scaling behaviours are actually possible ($\Delta_+=\Delta$ and $\Delta_-=d-\Delta$), corresponding to the two branches of solutions of the bulk Klein-Gordon equation.

\noindent\textit{Spin-two analogue:} The Ricci flatness of the ambient metric is the analogue of the harmonicity of the ambient scalar field, while Einstein equation with a cosmological constant of bulk gravity is the analogue of Klein-Gordon equation of the bulk scalar field. The holographic reconstruction of the $(d+1)$-dimensional spacetime metric from the $d$-dimensional conformal boundary data via the $(d+2)$-dimensional ambient space is the essence of FG construction. 

\paragraph{On-shell boundary scalar field (aka singleton).}

For generic values of the conformal dimension $\Delta$,
the harmonicity $\nabla^2 \Phi=0$ is incompatible with the gauge equivalence $\Phi\sim \Phi + V^2\, \lambda$\,.
In such cases, the harmonicity can be interpreted as a gauge-fixing condition for the gauge equivalence. 
This can be summarized by saying that the off-shell boundary scalar field can be equivalently described as the ambient scalar satisfying  \eqref{C1} or \eqref{C2}.
However, when the conformal dimension takes the value $\Delta=(d-2)/2$ one can impose consistently 
\begin{equation}\label{C3}
\nabla^2 \Phi=0\,,\qquad \left(V^A \d_A+\frac{d-2}{2}\right)\Phi=0\,, \qquad
\Phi\sim \Phi + V^2\,\lambda\,.
\end{equation}
The consistency of the system follows from  
the fact that (as was originally observed in~\cite{GJMS}) the operators $\nabla^2$, $V^A\d_A+\frac{d+2}{2}$, $V^2$ form the algebra $\mathfrak{sp}(2)$. Note that these operators can be thought as first-quantized versions of the $\mathfrak{sp}(2)$-constraints $F_+$, $F_0$, $F_-$.

Identifying an equivalence class determined by the second and the third relations in \eqref{C3} with an off-shell boundary scalar field,
the first equation imposes the Yamabe equation for this conformal scalar field in $d$ dimensions.
Another interpretation of the same fact is that the first two constraints describe the bulk  Klein-Gordon equation with critical mass. Considered in the vicinity of the boundary, the equivalence relation described by the third relation in~\eqref{C3} then eliminates the subleading solutions. From this perspective, the Yamabe equation appears as an obstruction in extending the unconstrained boundary value to
an on-shell bulk field (or, equivalently, as an obstruction in the near-boundary expansion of the on-shell bulk scalar field).

\noindent\textit{Spin-two analogue:} Consider $d=4$ for simplicity (but similar discussion holds for any even $d$). The Bach tensor appears as an obstruction (related to the holographic anomaly) in the FG expansion of bulk gravity in five dimensions. This obstruction is absent if and only if four-dimensional conformal gravity is on-shell.

%%%%%%%%%%%%%%%%%%%%%%%%%%%%%%%%%%%%%%%%%%%%%%%%%%%%%%%%%%%%%%%%
\end{appendix}
%%%%%%%%%%%%%%%%%%%%%%%%%%%%%%%%%%%%%%%%%%%%%%%%%%%%%%%%%%%%%%%%
\providecommand{\href}[2]{#2}\begingroup\raggedright\endgroup

\end{document}